\documentclass[aps,prb,twocolumn]{revtex4-1}

\usepackage[english]{babel}
\usepackage[utf8]{inputenc}
\usepackage{amsmath}
\usepackage{amssymb}
\usepackage[caption=false]{subfig}
\usepackage{amssymb}
\usepackage{epsfig}
\usepackage{graphicx}
\usepackage{amsmath}
\usepackage{array,color}
\usepackage{natbib}

\usepackage{soul}
\usepackage{textcomp}
\usepackage[T1]{fontenc}

\usepackage[usenames,dvipsnames]{xcolor}
\definecolor{forestgreen}{rgb}{0.11,0.54,0.15}
\definecolor{purple}{rgb}{0.62,0.10,0.96}
\definecolor{dockerblue}{rgb}{0.11,0.56,0.98}
\definecolor{freeblue}{rgb}{0.25,0.41,0.88}

\usepackage[pdftex,plainpages=false,colorlinks=true,linkcolor=Red, citecolor=blue, urlcolor=blue]{hyperref}

\begin{document}

\title{Magic DIAMOND: Multi-Fascicle Diffusion Compartment Imaging with Tensor Distribution Modeling and Tensor-Valued Diffusion Encoding}

\author{A. Reymbaut$^{1,2}$}
\email{alexis.reymbaut@usherbrooke.ca}
\author{A. Valcourt Caron$^1$, G. Gilbert$^{2}$, F. Szczepankiewicz$^{3,4}$, M. Nilsson$^3$, S. K. Warfield$^5$}
\author{M. Descoteaux$^{1}$}
\thanks{Co-last authors. These senior authors contributed equally to this work.}
\author{B. Scherrer$^{5}$}
\thanks{Co-last authors. These senior authors contributed equally to this work.}

\affiliation{
$^1$Universit\'{e} de Sherbrooke, Sherbrooke, QC  J1K 2R1, Canada\\
$^2$MR Clinical Science, Philips Healthcare Canada, Markham, ON L6C 2S3, Canada\\
$^3$Department of Clinical Sciences, Lund University, 22184, Lund, Sweden\\
$^4$Random Walk Imaging AB, 22224, Lund, Sweden\\
$^5$Department of Radiology, Boston Children's Hospital, Boston, MA 02115, United States
}

\date{\today}

\begin{abstract}
Diffusion tensor imaging provides increased sensitivity to microstructural tissue changes compared to conventional anatomical imaging but also presents limited specificity. To tackle this problem, the DIAMOND model subdivides the voxel content into diffusion compartments and draws from diffusion-weighted data to estimate compartmental non-central matrix-variate Gamma distribution of diffusion tensors, thereby resolving crossing fascicles while accounting for their respective heterogeneity. Alternatively, tensor-valued diffusion encoding defines new acquisition schemes tagging specific features of the intra-voxel diffusion tensor distribution directly from the outcome of the measurement. However, the impact of such schemes on estimating brain microstructural features has only been studied in a handful of parametric single-fascicle models. In this work, we derive a general Laplace transform for the non-central matrix-variate Gamma distribution, which enables the extension of DIAMOND to tensor-valued encoded data. We then evaluate this "Magic DIAMOND" model \textit{in silico} and \textit{in vivo} on various combinations of tensor-valued encoded data. Assessing uncertainty on parameter estimation \textit{via} stratified bootstrap, we investigate both voxel-based and fixel-based metrics by carrying out multi-peak tractography. We show that our estimated metrics can be mapped along tracks robustly across regions of fiber crossing, which opens new perspectives for tractometry and microstructure mapping along specific white-matter tracts. 
\end{abstract}

\maketitle

\textbf{Abbreviations used:} MRI, magnetic resonance imaging; DWI, diffusion-weighted MRI; DW, diffusion-weighted; DTI, diffusion tensor imaging; DIAMOND, distribution of anisotropic microstructural environments in diffusion compartment imaging; CHARMED, composite hindered and restricted model of diffusion; NODDI, neurite orientation dispersion and density imaging; ADC, apparent diffusion coefficient; FW, free-water; fMD, fascicle mean diffusivity; fAD, fascicle axial diffusivity; fRD, fascicle radial diffusivity; fFA, fascicle fractional anisotropy; PGSE, pulsed gradient spin echo; AIC, Akaike information criterion; SNR, signal-to-noise ratio; IQR, interquartile range; LL, linear-linear diffusion encoding combination; LP, linear-planar diffusion encoding combination; LS, linear-spherical diffusion encoding combination.


\section{Introduction}

\begin{figure}[h!]
\begin{center}
\includegraphics[width=20pc]{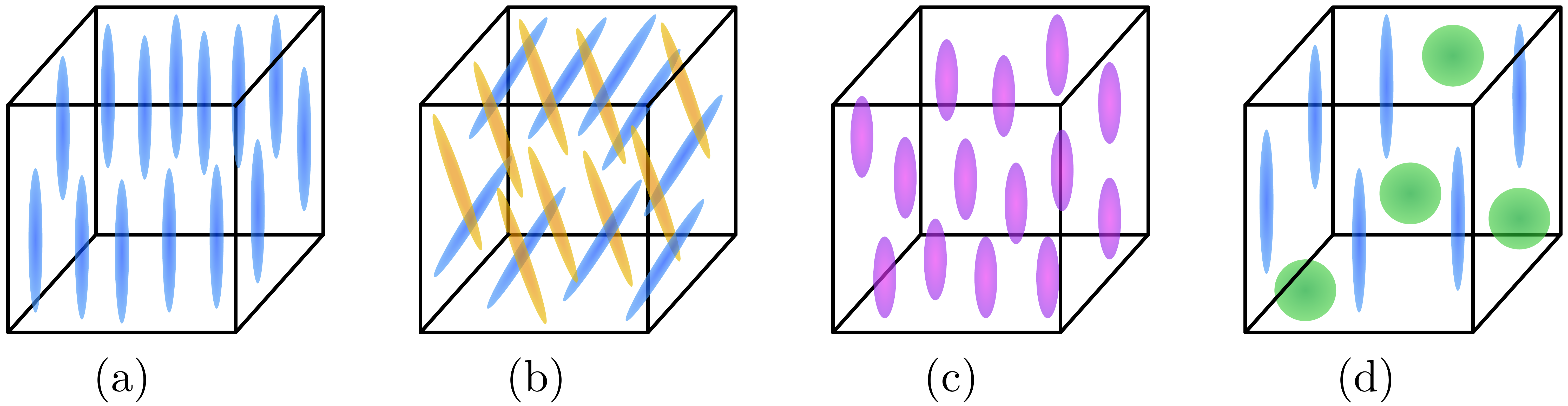}
\caption{Illustration of archetypal voxel contents in terms of colored microscopic diffusion tensors. While the voxel contents typically associated to a white-matter (WM) fiber (a) and to WM crossing fibers (b) would for instance correspond to the healthy corpus callosum and its crossing with the cingulum, respectively, the contents typically associated to demyelination (c) and WM inflammation (d) usually correspond to unhealthy white matter. DTI would typically yield comparable voxel-averaged diffusion tensors for the three voxels on the right.}
\label{Figure_voxels}
\end{center}
\end{figure}

Measuring water diffusion with diffusion-weighted MRI (DWI) enables the non-invasive characterization of biological tissues \textit{in vivo}. In particular, diffusion tensor imaging (DTI)~\citep{Basser:1994}, a very common DWI technique, has proven sensitive to microstructural tissue changes but only provides a voxel-scale average of the intra-voxel diffusion profile. Indeed, DTI considers the DW signal $\mathcal{S}(b,\mathbf{n})$ acquired during a typical Stejskal-Tanner sequence~\citep{Stejskal_Tanner:1965} with b-value $b$ and unit orientation $\mathbf{n}$ for the diffusion-probing magnetic field gradient, and interprets it as the following monoexponential:
\begin{equation}
\frac{\mathcal{S}(b,\mathbf{n})}{\mathcal{S}_0} = \exp(-b\,\mathbf{n}^\mathrm{T}\cdot \langle\mathbf{D}\rangle\cdot\mathbf{n}) = \exp(-bD_\mathbf{n}) \, ,
\label{Eq_signal_decay_DTI}
\end{equation}
where $\mathcal{S}_0$ is the non-DW signal, $\langle\mathbf{D}\rangle$ is the voxel-scale averaged diffusion tensor, $D_\mathbf{n}$ is the effective diffusivity along orientation $\mathbf{n}$, "$\cdot$" denotes the vector/tensor multiplication, and the superscript "$\mathrm{T}$" indicates vector/tensor transposition. As a result, DTI shows poor specificity in depicting the precise nature of microstructural tissue changes~\citep{Alexander:2001,Alexander:2007,Jones_Cercignani:2010,Jones:2012}, especially in voxels of crossing fascicles~\citep{Pierpaoli:1996,Douaud:2011}, which make up 60 to 90\% of the brain~\citep{Jeurissen:2013}. Fig.~\ref{Figure_voxels} illustrates a few archetypal voxels where DTI's lack of specificity may arise. 

DTI's lack of specificity stems from the inherent heterogeneity of typical DWI voxels, whose volume of a few cubic millimeters encompasses multiple cell types, sizes, geometries and orientations, and the extra-cellular space~\citep{Norris:2001,Sehy:2002,Minati:2007,Mulkern:2009}. In other words, the measured DW signal is a combination of signals arising from a variety of microstructural environments. Approximating that diffusion is a Gaussian process within these environments and that no exchange occurs between them~\citep{Johnson:1993}, which usually holds in healthy tissues for typical acquisition times~\citep{NilssonThesis:2011,Nilsson:2013a,Nilsson:2013b,Lampinen:2017}, intra-voxel heterogeneity can be accounted for by considering a weighted sum of all microstructural signals, thus obtaining the following DW signal:
\begin{equation}
\frac{\mathcal{S}(b,\mathbf{n})}{\mathcal{S}_0} = \int \mathcal{P}(\mathbf{D})\,  \exp(-b\,\mathbf{n}^\mathrm{T}\cdot \mathbf{D}\cdot\mathbf{n}) \, \mathrm{d}\mathbf{D} \, ,
\label{Eq_signal_decay_DTI_P}
\end{equation}
where the intra-voxel diffusion tensor distribution $\mathcal{P}(\mathbf{D})$ weighs the different microstructural contributions~\citep{Jian:2007} and the integral covers the space $\mathrm{Sym}^+(3)$ of 3$\times$3 symmetric positive-definite tensors. To better capture intra-voxel heterogeneity and thus overcome DTI's limitations, one can either attempt to improve the interpretation of the diffusion signal post-acquisition, \textit{via} an appropriate signal representation or model, or to enhance the specificity of the set of acquired signals itself. 

In this work, we combine two techniques that respectively target these two solutions: tensor-valued diffusion encoding~\citep{Cory:1990,Mori:1995,Wong:1995,Eriksson:2013,Westin:2014,Eriksson:2015,Westin:2016,Topgaard:2017} and the Distribution of anisotropic microstructural environments in diffusion compartment imaging (DIAMOND) model~\citep{Scherrer_DIAMOND:2016,Scherrer_aDIAMOND:2017}, that has shown great promise in mapping the Meyers loop~\citep{Chamberland:2017}, imaging the preterm human cortex~\citep{Eaton-Rosen:2017}, characterizing mild traumatic brain injuries~\citep{Scherrer_ISMRM:2017} and studying infantile autism~\citep{Scherrer:2019}. After introducing these techniques in sections~\ref{Sec_DIAMOND_model} and \ref{Sec_tensor_valued_diff}, we extend DIAMOND to tensor-valued encoded data in section~\ref{Sec_Magic_DIAMOND} and obtain the "Magic DIAMOND" model, named in homage to the magic-angle spinning DW acquisitions of Ref.~\onlinecite{Eriksson:2013} and its physical chemistry inspiration~\citep{Andrew:1959}. This new approach, first to draw from tensor-valued diffusion encoding to estimate compartmental distributions of diffusion tensors, is designed to tackle various problems such as edema, neuroinflammation or axonal degradation in voxels of crossing fascicles, all relevant to the majority of the brain and to the study and understanding of neurodegenerative diseases. In section~\ref{Sec_Methods}, we describe the methods used to evaluate Magic DIAMOND on \textit{in silico} signals and \textit{in vivo} human data. In particular, we address how to select the appropriate number of intra-voxel fascicles, estimate parameters, acquire the \textit{in silico} and \textit{in vivo} datasets, and compare our model with the original DIAMOND model on these datasets using stratified bootstrap (\textit{in silico} and \textit{in vivo}), cost function computation (\textit{in vivo}) and tractography (\textit{in vivo}). Results reporting on various combinations of diffusion encodings are presented in section~\ref{Sec_Results} and discussed in section~\ref{Sec_Discussion}. While the \textit{in silico} signals enable preliminary quantification of the accuracy and precision of Magic DIAMOND's estimations, the \textit{in vivo} data offers a proof of principle of Magic DIAMOND's potential.

\section{Theory}
\label{Sec_Theory}

\subsection{The DIAMOND model}
\label{Sec_DIAMOND_model}

Mathematical models enable a direct translation from the diffusion signal's features to metrics that should describe the biological properties of the voxel content, as discussed in~\citep{Yablonskiy:2010,Jelescu:2017,Novikov:2018}. Among them, diffusion compartment imaging (DCI) reflects the presence of intra-voxel tissue compartments and relates compartmental features to microstructural tissue properties at sub-voxel resolution, effectively providing insight into microstructural tissue changes with critically improved sensitivity and specificity. A brief review of major DCI techniques includes the Composite hindered and restricted model of diffusion (CHARMED)~\citep{Assaf_CHARMED:2004, Assaf_CHARMED:2005}, its extension AxCaliber~\citep{Assaf:2008}, and the Neurite orientation dispersion and density imaging (NODDI) model~\citep{Zhang_NODDI:2012}. However, these methods consider key assumptions that are inconsistent with the known tissue microstructure~\citep{Alexander:2010,Aboitiz:1992,Lampinen_CODIVIDE:2017,Nilsson:2017}. As a solution, fitting either compartment-specific distributions of apparent diffusion coefficients (ADCs)~\citep{Yablonskiy:2003} or one-dimensional Gamma distributions of ADCs in the ball-and-stick model~\citep{Jbabdi:2012} was proposed. However, while the former could not characterize the 3D anisotropy of diffusion observed in the brain~\citep{Moseley:1990}, the latter only implemented a Gamma distribution with same shape and scale parameters for all compartments. Finally, none of the previous models estimates crossing anisotropic compartments. 

\begin{figure}[h!]
\begin{center}
\includegraphics[width=20pc]{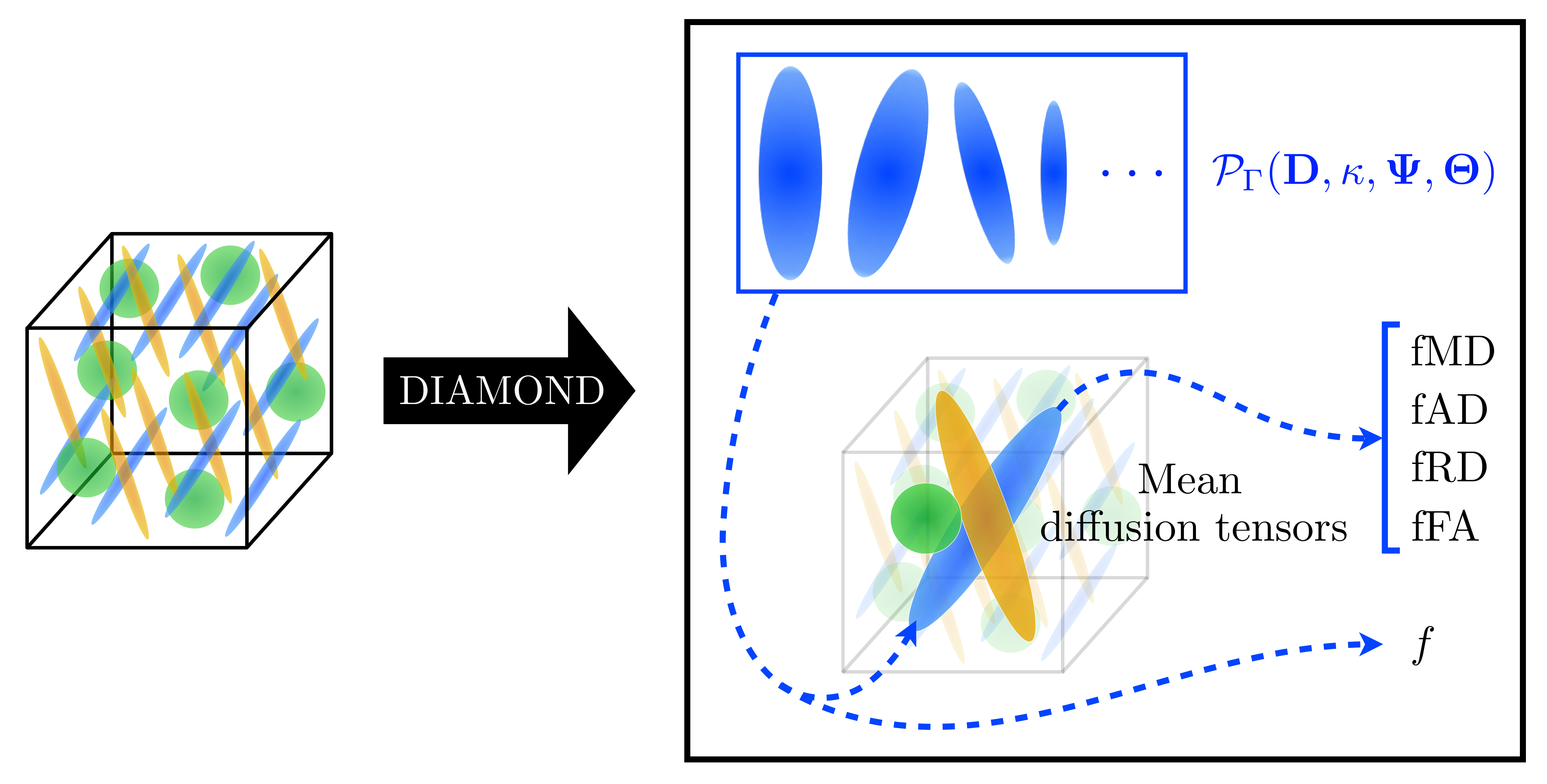}
\caption{Illustration of the way a complex voxel content (crossing fibers and free water, left panel) is depicted withing the DIAMOND model (right panel). Each anisotropic diffusion compartment is described in terms of a non-central matrix-variate Gamma distribution $\mathcal{P}_\Gamma (\mathbf{D},\kappa,\mathbf{\Psi},\mathbf{\Theta})$ Eq.~\eqref{Eq_mv_Gamma_distribution_asym} whose mean diffusion tensor is associated to various fascicle metrics, such as the fascicle mean diffusivity (fMD), the fascicle axial diffusivity (fAD), the fascicle radial diffusivity (fRD) and the fascicle fractional anisotropy (fFA). Each overall distribution is also given a total signal fraction $f$.}
\label{Figure_voxel_DIAMOND}
\end{center}
\end{figure}

A recent DCI model has been proposed to circumvent these limitations: the Distribution of anisotropic microstructural environments in diffusion compartment imaging (DIAMOND) model~\citep{Scherrer_DIAMOND:2016,Scherrer_aDIAMOND:2017}. Illustrated in Fig.~\ref{Figure_voxel_DIAMOND}, DIAMOND considers a free-water (FW) compartment and a series of anisotropic compartments per voxel. In the brain, these anisotropic compartments are usually referred to as "fascicles," as anisotropic diffusion is conventionally recognized to arise from the presence of white matter fascicles at the mesoscale. The signal arising from each fascicle is described by a distinct unimodal non-central matrix-variate Gamma distribution $\mathcal{P}_\Gamma (\mathbf{D},\kappa,\mathbf{\Psi},\mathbf{\Theta})$ of 3$\times$3 symmetric positive-definite diffusion tensors~\citep{Gupta_Nagar_Book:2000}:
\begin{widetext}
\begin{align}
\mathcal{P}_\Gamma (\mathbf{D},\kappa,\mathbf{\Psi},\mathbf{\Theta}) =& \;\frac{\mathrm{Det}(\mathbf{D})^{\kappa - 2}}{\mathrm{Det}(\mathbf{\Psi})^\kappa \, \Gamma_3(\kappa)}\, \exp(-\mathrm{Tr}(\mathbf{\Theta} +\mathbf{\Psi}^{-1} \cdot\mathbf{D}))\times\mathcal{F}_{0,1}(\kappa, \mathbf{\Theta}\cdot\mathbf{\Psi}^{-1}\cdot\mathbf{D})
\; ,
\label{Eq_mv_Gamma_distribution_asym}
\end{align}
\end{widetext}
where $\Gamma_3(\kappa)= \pi^{3/2}\prod_{m=1}^3 \Gamma(\kappa - (m-1)/2)$ is the multivariate Gamma function, $\mathcal{F}_{0,1}$ is the hypergeometric (Bessel) function of matrix argument of order $(0,1)$, $\kappa>1$ is the shape parameter, $\mathbf{\Psi}\in\mathrm{Sym}^+(3)$ is the scale tensor and $\mathbf{\Theta}\in\mathrm{Sym}(3)$ is the noncentrality parameter ($\mathrm{Sym}(3)$ denotes the space of 3$\times$3 symmetric tensors). Consequently, the overall modeled signal writes
\begin{equation}
\frac{\tilde{\mathcal{S}}(b,\mathbf{n})}{\mathcal{S}_0} = f_\mathrm{FW} \, \exp(-bD_\mathrm{FW}) + \sum_j f_j \, \frac{\tilde{\mathcal{S}}_{\mathrm{f},j}(b,\mathbf{n})}{\mathcal{S}_0} \, ,
\label{Eq_signal_decay_DTI_P_DIAMOND}
\end{equation}
with
\begin{equation}
\frac{\tilde{\mathcal{S}}_{\mathrm{f},j}(b,\mathbf{n})}{\mathcal{S}_0} = \int\mathcal{P}_\Gamma (\mathbf{D},\kappa_j,\mathbf{\Psi}_j,\mathbf{\Theta}_j)\, \exp(-b\,\mathbf{n}^\mathrm{T}\cdot\mathbf{D}\cdot\mathbf{n}) \, \mathrm{d}\mathbf{D}\, ,
\end{equation}
where $f_\mathrm{FW},f_j \in [0,1]$ are compartmental signal fractions, normalized so that $f_\mathrm{FW}+\sum_j f_j = 1$, and $D_\mathrm{FW} = 3$~\textmu$\mathrm{m}^2/\mathrm{ms}$. 

The mean diffusion tensor of the distribution is given as a function of the distribution parameters by
\begin{equation}
\langle\mathbf{D}\rangle = \mathbf{\Psi}\cdot(\kappa\mathbf{I}_3 + \mathbf{\Theta})\, .
\label{Eq_Psi}
\end{equation}
The "width" of the peak-shaped tensor distribution is interpreted in terms of "tissue heterogeneity", since a very heterogeneous compartment should be described using a variety of distinct diffusion tensors. While the shape parameter $\kappa$ relates to the isotropic component of tissue heterogeneity, the noncentrality parameter $\mathbf{\Theta}$ allows to squeeze the distribution in order to capture anisotropic tissue heterogeneity. Indeed, this effect of $\mathbf{\Theta}$ on any fit of Eq.~\eqref{Eq_signal_decay_DTI_P_DIAMOND} is ensured by the choice of parametrization made for $\mathbf{\Theta}$ in Ref.~\onlinecite{Scherrer_aDIAMOND:2017}: if one expresses $\langle\mathbf{D}\rangle$ and $\mathbf{\Theta}$ using the transfer matrix $\mathbf{V}$ from the laboratory frame of reference to the compartment's eigenbasis and to insert this result into Eq.~\eqref{Eq_Psi}, one then obtains
\begin{equation}
\langle\mathbf{D}\rangle = \mathbf{V}\cdot \mathrm{Diag}(\lambda^\perp,\lambda^\perp,\lambda^\parallel)\cdot \mathbf{V}^\mathrm{T}\, ,
\end{equation}
where $\lambda^\parallel \geq \lambda^\perp$, and 
\begin{equation}
\mathbf{\Theta} = \mathbf{V}\cdot \mathrm{Diag}(0,0,\kappa^\prime)\cdot \mathbf{V}^\mathrm{T}\, .
\label{Eq_Theta}
\end{equation}
Note that the FW signal in Eq.~\eqref{Eq_signal_decay_DTI_P_DIAMOND} corresponds to a non-central matrix-variate Gamma distribution with $\kappa \to +\infty$, $\kappa^\prime \to +\infty$ and isotropic $\langle\mathbf{D}\rangle$.

DIAMOND thus effectively allows the evaluation of compartment-specific diffusion features, such as fascicle mean diffusivity (fMD), fascicle axial diffusivity (fAD), fascicle radial diffusivity (fRD) and fascicle fractional anisotropy (fFA), while also accounting for intra-compartment heterogeneity, especially in voxels of crossing fascicles. However, extracting compartment-specific features from conventional DWI data is known to be a difficult inversion problem, involving functionals with numerous local minima. Besides, even if intra-compartment heterogeneity is accounted for by the use of a compartmental diffusion tensor distribution - a mathematical achievement in itself - properly quantifying this heterogeneity remains a challenge to this day. Indeed, analytically tractable definitions for such heterogeneity metrics have, to our knowledge, not yet been derived from parametric tensor distributions.

\subsection{Tensor-valued diffusion encoding}
\label{Sec_tensor_valued_diff}

An alternative way to enhance the specificity of the information provided by diffusion MRI is to obtain complementary pieces of diffusion information at the acquisition stage. The most commonly used diffusion sequence, the pulsed gradient spin echo (PGSE) sequence~\citep{Stejskal_Tanner:1965}, corresponds to applying a linear (unidirectional) diffusion gradient $\mathbf{g}(t)=g(t)\,\mathbf{n}$. Such a sequence is characterized by a spin-dephasing vector (or q-vector) $\mathbf{q}(t) = \gamma \int_0^t \mathbf{g}(t')\, \mathrm{d}t' = q(t)\, \mathbf{n}$ that grows and vanishes through time along a \textit{unique} direction given by the unit vector $\mathbf{n}$. The time-dependent norm of this q-vector yields the b-value $b = \int_0^\tau q^2(t)\, \mathrm{d}t$, where $\tau$ is the total encoding time. Instead, tensor-valued diffusion encoding (or b-tensor encoding)~\citep{Westin:2014,Eriksson:2015,Westin:2016,Topgaard:2017} has introduced q-trajectories in which the q-vector $\mathbf{q}(t) = \gamma \int_0^t \mathbf{g}(t')\, \mathrm{d}t' = q(t)\, \mathbf{n}(t)$ follows a continuous and non-trivial trajectory through time~\citep{Eriksson:2013,Westin:2016}. The name "b-tensor encoding" comes from the fact that the time-dependency of the q-vector's orientation $\mathbf{n}(t)$ does not encode diffusion weighting in a vector, but rather in a 3$\times$3 symmetric positive-semidefinite tensor, 
\begin{equation}
\mathbf{b} = \int_0^\tau q^2(t)\, \mathbf{n}(t)\cdot \mathbf{n}^\mathrm{T}(t)\,\mathrm{d}t \in \mathrm{Sym}^{(+)}(3) \, ,
\label{Eq_b_tensor}
\end{equation}
where $\mathrm{Sym}^{(+)}(3)$ denotes the space of symmetric positive-semidefinite tensors. Note that this b-tensor was introduced in earlier works as a b-matrix accounting for cross terms between diffusion gradient pulses~\citep{Basser:1994,Mattiello:1997}. 

While the shape of the b-tensor's glyph gives its name to the corresponding diffusion encoding (linear or spherical encoding for instance), its trace defines the associated b-value $b = \mathrm{Tr}(\mathbf{b})$. In this general b-tensor formalism, the specific tagging of a given diffusion pattern is achieved through the expression of the overall DW signal \citep{Westin:2014,Westin:2016}:
\begin{equation}
\frac{\mathcal{S}(\mathbf{b})}{\mathcal{S}_0} = \int\mathcal{P}(\mathbf{D})\,  \exp(-\mathbf{b}:\mathbf{D}) \, \mathrm{d}\mathbf{D} \; ,
\label{Eq_signal_decay_DTI_P_b-tensor}
\end{equation}
where
\begin{equation}
\mathbf{b}:\mathbf{D} = \sum_{ij} b_{ij} D_{ij} = \int_0^\tau q^2(t)\;\mathbf{n}^{\mathrm{T}}(t)\cdot \mathbf{D}\cdot \mathbf{n}(t)\, \mathrm{d}t
\label{Eq_Frobenius}
\end{equation}
is the Frobenius inner product, a generalized scalar product that "projects" the diffusion tensor onto the b-tensor, thereby effectively selecting a certain diffusion pattern to probe. For instance, the signal Eq.~\eqref{Eq_signal_decay_DTI_P}, associated to the PGSE q-vector $\mathbf{q}(t) = q(t)\, \mathbf{n}$, actually maps back to the linear (stick-like) b-tensor $\mathbf{b} = b \;\mathbf{n}\cdot\mathbf{n}^\mathrm{T}$ that was already introduced in~\citep{Jian:2007}. Also, the spherical encoding yielded by a spherical b-tensor allows for the specific encoding of isotropic diffusion during the acquisition process~\citep{Mori:1995,Wong:1995,Eriksson:2013}. Combinations of these two encodings, coupled to the fitting of a Gamma distribution of ADCs on powder-averaged data, have already enabled the extraction of novel diffusion metrics~\citep{Lasic:2014,Eriksson:2015} that may be used to measure the microscopic diffusion anisotropy of healthy brain tissues~\citep{Szczepankiewicz:2015}, as well as to probe diffusional variance in tumors~\citep{Szczepankiewicz:2016}.

Even though the opportunity to measure non-conventional pieces of diffusion information dates back almost 30 years, with the double diffusion encoding of Ref.~\onlinecite{Cory:1990} and the triple diffusion encoding of Refs.~\onlinecite{Mori:1995,Wong:1995}, it is only within the last decade that non-trivial diffusion encoding has drawn much attention~\citep{Ozarslan:2009,Shemesh:2010,Lawrenz:2010,Valette:2012,Jespersen:2012,Jespersen:2013,Lawrenz:2016,Ianus:2016,Lawrenz:2018,Ianus:2018,Jensen_Helpern:2018,Szczepankiewicz_Jeurissen_ISMRM:2018,Herberthson:2019}. From a modeling standpoint, the main interest lies in assessing whether the estimation of microstructural features benefits from these non-trivial diffusion encodings or not. The impact of planar and spherical encodings has already been investigated in ball-and-stick-like models such as the white matter Standard Model~\citep{Novikov_WMSM:2018} and NODDI~\citep{Zhang_NODDI:2012}, focusing mainly on accuracy~\citep{Lampinen_CODIVIDE:2017}, precision~\citep{Fieremans_ISMRM:2018} and degeneracy~\citep{Reisert:2019,Coelho:2019,Coelho_arxiv:2019} in parameter estimation. Also, a generalized cumulant approach implicitly assuming a normal distribution of diffusion tensors has already been implemented in the framework of b-tensor encoding~\citep{Westin:2016}. However, these models do not separately characterize crossing fascicles. One exception lies in the Monte-Carlo inversion of Refs.~\onlinecite{deAlmeidaMartins:2016,deAlmeidaMartins:2018,Topgaard:2019} that shows impressive results in teasing apart intra-voxel populations in the diffusion-relaxation space but can unfortunately be significantly time-consuming and noise-sensitive. 

\begin{widetext}

\subsection{The Magic DIAMOND model}
\label{Sec_Magic_DIAMOND}

Extending the DIAMOND model to tensor-valued encoded diffusion data, a procedure illustrated in Fig.~\ref{Figure_Scheme}, consists in merging Eqs.~\eqref{Eq_signal_decay_DTI_P_DIAMOND} and \eqref{Eq_signal_decay_DTI_P_b-tensor} to yield the following Magic DIAMOND signal:
\begin{align}
\frac{\tilde{\mathcal{S}}(\mathbf{b})}{\mathcal{S}_0} = &\; f_\mathrm{FW} \, \exp(-bD_\mathrm{FW}) + \sum_j f_j \underbrace{\int\mathcal{P}_\Gamma (\mathbf{D},\kappa_j,\mathbf{\Psi}_j,\mathbf{\Theta}_j)\,  \exp(-\mathbf{b}:\mathbf{D}) \, \mathrm{d}\mathbf{D}}_{\tilde{\mathcal{S}}_{\mathrm{f},j}(\mathbf{b})/\mathcal{S}_0}  \, ,
\label{Eq_signal_decay_DTI_P_Magic_DIAMOND}
\end{align}
whose fascicle part $\tilde{\mathcal{S}}_{\mathrm{f},j}(\mathbf{b})/\mathcal{S}_0$ remains to be analytically computed. This fascicle part is the general Laplace transform of the non-central matrix-variate Gamma distribution Eq.~\eqref{Eq_mv_Gamma_distribution_asym}. 

\begin{figure*}[h!]
\begin{center}
\includegraphics[width=25pc]{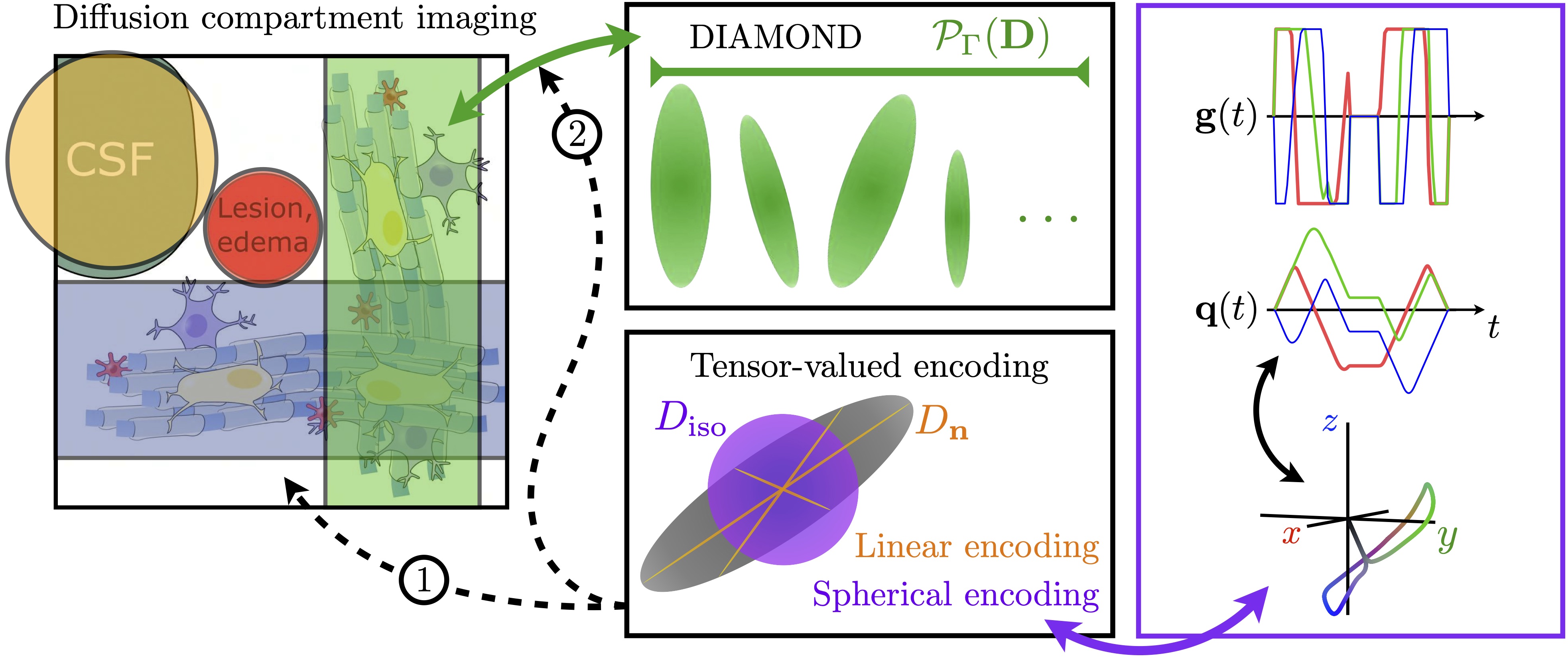}
\caption{Diffusion compartement imaging (DCI) techniques subdivide the voxel content into diffusion compartments (left panel). In particular, the DIAMOND model estimates a non-central matrix-variate Gamma distribution of diffusion tensors $\mathcal{P}_\Gamma(\mathbf{D})$ for each compartment (top center panel). Alternatively, tensor-valued encoding (bottom center panel) establishes novel sequences that yield for instance spherical encoding, which specifically tags isotropic diffusion patterns ($D_\mathrm{iso}$, in purple) of any diffusion tensor (grey glyph), contrary to the conventional linear encoding that only probes diffusion along single directions $\mathbf{n}$ ($D_{\mathbf{n}}$, in orange). The spherically encoded sequence used throughout this paper (gradient waveform $\mathbf{g}(t)$, spin-dephasing vector $\mathbf{q}(t)$ and q-trajectory) is presented as an example in the right panel, where color codes for orientation. In this work, we incorporate tensor-valued encoding in the DIAMOND fitting procedure at two key steps (dashed arrow lines): 1) during the acquisition process, 2) in the estimation of $\mathcal{P}_\Gamma(\mathbf{D})$ for each compartment.}
\label{Figure_Scheme}
\end{center}
\end{figure*}

\subsubsection{General Laplace transform of the non-central matrix-variate Gamma distribution} 

Let us define the general Laplace transform of an arbitrary diffusion tensor distribution $\mathcal{P}(\mathbf{D})$ from Eq.~\eqref{Eq_signal_decay_DTI_P_b-tensor} as
\begin{equation}
\mathcal{T}_\mathcal{P}(\mathbf{b}) = \int\mathcal{P}(\mathbf{D})\,  \exp(-\mathbf{b}:\mathbf{D}) \, \mathrm{d}\mathbf{D}\, ,
\label{Eq_Laplace_transform_general}
\end{equation}
where $\mathcal{T}$ denotes the Laplace transformation. To compute this Laplace transform, let us introduce the moment-generating function for any distribution $\mathcal{P}$ of symmetric tensors~\citep{Gupta_Nagar_Book:2000}:
\begin{equation}
M_\mathcal{P}(\mathbf{Z}) = \int \! \mathcal{P}(\mathbf{D})\, \exp\!\left(\mathrm{Tr}[\mathbf{Z} \cdot \mathbf{D}]\right)  \mathrm{d}\mathbf{D}\, ,
\label{Eq_moment_generating_function}
\end{equation}
where $\mathbf{Z} \in \mathrm{Sym}(3)$. By choosing the symmetric tensor
\begin{equation}
\mathbf{Z} \equiv -\mathbf{b} = -\int_0^\tau\! q^2(t) \, \mathbf{n}(t) \cdot \mathbf{n}^\mathrm{T}(t)\, \mathrm{d}t
\label{Eq_New_choice}
\end{equation}
from Eq.~\eqref{Eq_b_tensor} and by using the trace property $\mathrm{Tr}(\mathbf{u}\cdot\mathbf{v}^\mathrm{T}) = \mathbf{v}^\mathrm{T} \cdot \mathbf{u}$ and the definition of the Frobenius inner product Eq.~\eqref{Eq_Frobenius}, we prove that the Laplace transform Eq.~\eqref{Eq_Laplace_transform_general} equals 
\begin{equation}
\mathcal{T}_\mathcal{P}(\mathbf{b}) = M_\mathcal{P}(-\mathbf{b}) \, .
\label{Eq_Laplace_moment_generating_function}
\end{equation}
In the particular case of the non-central matrix-variate Gamma distribution $\mathcal{P}_\Gamma$ Eq.~\eqref{Eq_mv_Gamma_distribution_asym}, it can be shown~\citep{Gupta_Nagar_Book:2000} that its moment-generating function is given by
\begin{equation}
M_{\mathcal{P}_\Gamma}(\mathbf{Z}) = [\mathrm{Det}(\mathbf{I}_3 - \mathbf{Z}\cdot\mathbf{\Psi})]^{-\kappa} \; \mathrm{exp}\!\left[\mathrm{Tr}([(\mathbf{I}_3 - \mathbf{Z}\cdot\mathbf{\Psi})^{-1} - \mathbf{I}_3]\cdot \mathbf{\Theta}) \right] ,
\label{Eq_DIAMOND_Z}
\end{equation}
with $\mathbf{Z}$ such that $(\mathbf{I}_3 - \mathbf{Z}\cdot \mathbf{\Psi}) \in \mathrm{Sym}^+(3)$. Since the choice $\mathbf{Z} \equiv -\mathbf{b} $ Eq.~\eqref{Eq_New_choice} satisfies this last condition, the general Laplace transform of the non-central matrix-variate Gamma distribution writes
\begin{align}
\mathcal{T}_{\mathcal{P}_\Gamma}(\mathbf{b}) & = \int\mathcal{P}_\Gamma (\mathbf{D},\kappa,\mathbf{\Psi},\mathbf{\Theta})\,  \exp(-\mathbf{b}:\mathbf{D}) \, \mathrm{d}\mathbf{D} \nonumber \\
& = M_{\mathcal{P}_\Gamma}(-\mathbf{b})\nonumber \\
 & = [\mathrm{Det}(\mathbf{I}_3 + \mathbf{b} \cdot\mathbf{\Psi})]^{-\kappa} \; \mathrm{exp}\!\left[\mathrm{Tr}([(\mathbf{I}_3 + \mathbf{b}\cdot\mathbf{\Psi})^{-1} - \mathbf{I}_3]\cdot \mathbf{\Theta}) \right] .
 \label{Eq_Laplace_1}
\end{align}
This expression for $\mathcal{T}_{\mathcal{P}_\Gamma}(\mathbf{b})$ can be rewritten by considering the Woodbury matrix identity to simplify the previous trace: given a square invertible $n\times n$ matrix $\mathbf{A}$, an $n\times m$ matrix $\mathbf{U}$ and an $m\times n$ matrix $\mathbf{V}$, assuming that $(\mathbf{I}_m + \mathbf{V}\cdot\mathbf{A}^{-1}\cdot \mathbf{U})$ is invertible, one has $(\mathbf{A} + \mathbf{U}\cdot\mathbf{V})^{-1} = \mathbf{A}^{-1} - \mathbf{A}^{-1}\cdot \mathbf{U}\cdot (\mathbf{I}_m + \mathbf{V}\cdot\mathbf{A}^{-1}\cdot \mathbf{U})^{-1} \cdot \mathbf{V}\cdot \mathbf{A}^{-1}$. Taking $\mathbf{A}\equiv \mathbf{I}_3$, $\mathbf{U}\equiv \mathbf{b}$ and $\mathbf{V} \equiv \mathbf{\Psi}$, we yield
\begin{equation}
\mathcal{T}_{\mathcal{P}_\Gamma}(\mathbf{b})  = [\mathrm{Det}(\mathbf{I}_3 + \mathbf{b} \cdot\mathbf{\Psi})]^{-\kappa} \; \mathrm{exp}\!\left[-\mathbf{b} :  [(\mathbf{I}_3 + \mathbf{\Psi}\cdot\mathbf{b})^{-1} \cdot \mathbf{\Psi}\cdot \mathbf{\Theta} ] \right] ,
\label{Eq_Laplace_2}
\end{equation}
which is valid for any arbitrary b-tensor.

\subsubsection{Diffusion signal within the Magic DIAMOND model} 

Returning to Eq.~\eqref{Eq_signal_decay_DTI_P_Magic_DIAMOND}, using Eqs.~\eqref{Eq_Laplace_1} and \eqref{Eq_Laplace_2}, and omitting the compartmental index $j$, we obtain the Magic DIAMOND fascicle signal
\begin{align}
\frac{\tilde{\mathcal{S}}_\mathrm{f}(\mathbf{b})}{\mathcal{S}_0} 
 & =[\mathrm{Det}(\mathbf{I}_3 + \mathbf{b} \cdot\mathbf{\Psi})]^{-\kappa} \; \mathrm{exp}\!\left[\mathrm{Tr}([(\mathbf{I}_3 + \mathbf{b}\cdot\mathbf{\Psi})^{-1} - \mathbf{I}_3]\cdot \mathbf{\Theta}) \right] \nonumber \\
 & = [\mathrm{Det}(\mathbf{I}_3 + \mathbf{b} \cdot\mathbf{\Psi})]^{-\kappa} \; \mathrm{exp}\!\left[-\mathbf{b} :  [(\mathbf{I}_3 + \mathbf{\Psi}\cdot\mathbf{b})^{-1} \cdot \mathbf{\Psi}\cdot \mathbf{\Theta} ] \right]
\end{align}
for any arbitrary b-tensor. Let us render that signal expression more explicit in the case of an axisymmetric b-tensor. Using the previous expressions Eqs.~\eqref{Eq_Psi} and \eqref{Eq_Theta} for $\mathbf{\Psi}$ and $\mathbf{\Theta}$, the Magic DIAMOND fascicle signal writes
\begin{align}
\frac{\tilde{\mathcal{S}}_\mathrm{f}(b_\mathrm{L}, b_\mathrm{S},\beta)}{\mathcal{S}_0} = \left[ f_\perp(b_\mathrm{S})\,F(b_\mathrm{L}, b_\mathrm{S},\beta) \right]^{-\kappa} \mathrm{exp}\!\left(-\kappa^\prime\left[ 1 +\frac{f(b_\mathrm{L}, b_\mathrm{S},\beta)\,[1-f_\perp(b_\mathrm{S})]-1}{F(b_\mathrm{L}, b_\mathrm{S},\beta)} \right] \right) ,
 \label{Eq_Magic-DIAMOND_general} 
\end{align}
with the dimensionless quantities
\begin{align}
f_{\parallel,\perp}(b_\mathrm{S}) = &\; 1 + \frac{b_\mathrm{S}}{3}\, \frac{\lambda^{\parallel,\perp}}{\kappa^{\parallel,\perp}}\, , \\
f(b_\mathrm{L}, b_\mathrm{S}, \beta) = &\; 1 + \frac{3b_\mathrm{L}}{b_\mathrm{S}}\,\sin^2\beta\, , \\
F(b_\mathrm{L}, b_\mathrm{S}, \beta) = &\; f_{\parallel}(b_\mathrm{S})\,f_{\perp}(b_\mathrm{S}) + b_\mathrm{L} \left( \frac{\lambda^\parallel}{\kappa^\parallel}\cos^2\beta + \frac{\lambda^\perp}{\kappa^\perp}\sin^2\beta + \frac{b_\mathrm{S}}{3}\,\frac{\lambda^\parallel\lambda^\perp}{\kappa^\parallel\kappa^\perp}\right) .
\end{align}
Here, $\kappa^\parallel = \kappa + \kappa^\prime$ and $\kappa^\perp = \kappa$ are the axial and radial shape parameters, respectively, $\beta$ is the angle separating the axes of revolution of the diffusion and b- tensors, and $b_\mathrm{L}=bb_\Delta$ and $b_\mathrm{S}=b(1-b_\Delta)$ are two shape parameters drawn from Ref.~\onlinecite{Westin:2002} for the b-tensor. These parameters both depends on the normalized anisotropy parameter $b_\Delta\in[-0.5,1]$ introduced in Ref.~\onlinecite{Haeberlen:1976}, so that $\tilde{\mathcal{S}}_\mathrm{f}(b_\mathrm{L}, b_\mathrm{S},\beta)\equiv \tilde{\mathcal{S}}_\mathrm{f}(b, b_\Delta,\beta)$. In particular, $b_\Delta=1$ yields the linear signal
\begin{equation}
\frac{\tilde{\mathcal{S}}_\mathrm{f}^\mathrm{(lin.)}(b,\beta)}{\mathcal{S}_0} = \left[ 1 + b \left( \frac{\lambda^\parallel}{\kappa^\parallel}\cos^2\beta + \frac{\lambda^\perp}{\kappa^\perp}\sin^2\beta \right) \right]^{-\kappa^\perp}  \mathrm{exp}\!\left[\frac{ \displaystyle -b\, (\kappa^\parallel - \kappa^\perp)\,\lambda^\parallel \cos^2\beta}{\displaystyle \kappa^\parallel+ b \left( \lambda^\parallel \cos^2\beta + \frac{\kappa^\parallel}{\kappa^\perp}\,\lambda^\perp\sin^2\beta \right)} \right] ,
\label{Eq_Magic-DIAMOND_linear}
\end{equation}
$b_\Delta=0$ yields the spherical signal
\begin{equation}
\frac{\tilde{\mathcal{S}}_\mathrm{f}^\mathrm{(sph.)}(b)}{\mathcal{S}_0} = \left[ \left( 1 + \frac{b}{3}\,\frac{\lambda^\perp}{\kappa^\perp}\right)^2 \left(1 + \frac{b}{3}\,\frac{\lambda^\parallel}{\kappa^\parallel} \right) \right]^{-\kappa^\perp} \mathrm{exp}\!\left[ \frac{-b\, (\kappa^\parallel - \kappa^\perp)\, \lambda^\parallel}{3\kappa^\parallel + b\,\lambda^\parallel} \right] ,
\label{Eq_Magic-DIAMOND_spherical}
\end{equation}
and $b_\Delta=-0.5$ yields the planar signal
\begin{equation}
\frac{\tilde{\mathcal{S}}_\mathrm{f}^\mathrm{(plan.)}(b,\beta)}{\mathcal{S}_0} = \left( 1+ \frac{b}{2}\, \frac{\lambda^\perp}{\kappa^\perp} \right)^{-\kappa^\perp} \left[ 1 + \frac{b}{2} \left( \frac{\lambda^\parallel}{\kappa^\parallel}\sin^2\beta + \frac{\lambda^\perp}{\kappa^\perp}\cos^2\beta \right) \right]^{-\kappa^\perp} \mathrm{exp}\!\left[\frac{ \displaystyle - b\, 
(\kappa^\parallel - \kappa^\perp)\, \lambda^\parallel \sin^2\beta}{\displaystyle 2\kappa^\parallel + b \left( \lambda^\parallel \sin^2\beta + \frac{\kappa^\parallel}{\kappa^\perp}\, \lambda^\perp \cos^2\beta \right) } \right] .
\label{Eq_Magic-DIAMOND_planar}
\end{equation}
This new set of equations satisfies two key conditions. First, the linear signal Eq.~\eqref{Eq_Magic-DIAMOND_linear} directly translates back to the original DIAMOND signal of Ref.~\onlinecite{Scherrer_aDIAMOND:2017}. Second, this set of equations obeys the relationship linking linear, planar and spherical diffusion encodings for a purely homogeneous fascicle ($\kappa \to +\infty$ and $\kappa^\prime \to +\infty$):
\begin{equation}
\lim\limits_{\substack{\kappa^\prime \to +\infty\\\kappa \to +\infty}}\left[\tilde{\mathcal{S}}_\mathrm{f}^\mathrm{(lin.)}\left(\frac{b}{2},\beta\right)\,\tilde{\mathcal{S}}_\mathrm{f}^\mathrm{(plan.)}\left(b,\beta\right)\right] = \mathcal{S}_0 \lim\limits_{\substack{\kappa^\prime \to +\infty\\\kappa \to +\infty}}\left[\tilde{\mathcal{S}}_\mathrm{f}^\mathrm{(sph.)}\left(\frac{3b}{2}\right)\right] .
\end{equation}
\end{widetext}

\vspace*{\fill}

\newpage 
\clearpage
\newpage

\section{Material and methods}
\label{Sec_Methods}

\subsection{Model selection} 

As shown in Eq.~\eqref{Eq_signal_decay_DTI_P_Magic_DIAMOND}, Magic DIAMOND considers at each voxel a single free water (FW) compartment and a series of $N_\mathrm{f}$ fascicles (anisotropic compartments):
\begin{equation}
\frac{\tilde{\mathcal{S}}(\mathbf{b})}{\mathcal{S}_0} = f_\mathrm{FW} \, \exp(-bD_\mathrm{FW}) + \sum_j^{N_\mathrm{f}} f_j \, \frac{\tilde{\mathcal{S}}_{\mathrm{f},j}(\mathbf{b})}{\mathcal{S}_0} \, ,
\label{Eq_fractions}
\end{equation}
with $\tilde{\mathcal{S}}_{\mathrm{f},j}(\mathbf{b})/\mathcal{S}_0$ given by Eq.~\eqref{Eq_Magic-DIAMOND_general}. Notice that with this choice, intra-axonal restricted diffusion and extra-axonal hindered diffusion arising from a fascicle are both modeled using one tensor distribution. In order to set the number of intra-voxel fascicles ($N_\mathrm{f} \in \{0,1,2,3\}$ in this work), model selection was performed using the Akaike information criterion (AIC)~\citep{Akaike:1974}, promoting the model of lowest index
\begin{equation}
\mathrm{AIC} = 2n_\mathrm{param} - 2\ln(\mathcal{L})\, ,
\end{equation}
with $n_\mathrm{param}$ the number of estimated model parameters and $\mathcal{L}$ the maximum value of the model likelihood function. This criterion favors the goodness of fit while discouraging overfitting, but is not entirely impervious to noise in the data~\citep{Novikov:2018}. To ensure a fast and robust model selection, the Akaike model selection was performed on ball-and-stick models.

\subsection{Parameter estimation} 
\label{Sec_parameter_estimation}

The parameters of any evaluated diffusion tensor distribution were estimated within each voxel using a maximum \textit{a posteriori} approach where no prior correlation between the expectations $\langle\mathbf{D}\rangle$, fractions $\mathbf{f}$ and shape parameters $\bm{\kappa}$ is assumed. To ensure positive-definiteness of the diffusion tensors, we parametrized them using $\mathbf{L} \equiv \ln(\langle\mathbf{D}\rangle)$. The optimal map of parameters was then obtained through the Bayes maximization
\begin{widetext}
\begin{equation}
\left\{ \mathbf{L}_\mathrm{opt}, \bm{\kappa}_\mathrm{opt}, \mathbf{f}_\mathrm{opt} \right\} = \underset{\mathbf{L}, \bm{\kappa}, \mathbf{f}}{\mathrm{argmax}}\left[ P (\mathbf{I} \vert \mathbf{L},\bm{\kappa}, \mathbf{f})\, P (\mathbf{f} \vert \mathbf{L},\bm{\kappa})\, P (\bm{\kappa} \vert \mathbf{L})\, P (\mathbf{L}) \right] ,
\end{equation}
\end{widetext}
where $\mathbf{I}$ denotes the set of acquired images. While the probability $P (\mathbf{I} \vert \mathbf{L},\bm{\kappa}, \mathbf{f})$ was assumed to obey a normal distribution, 
$P(\mathbf{f} \vert \mathbf{L},\bm{\kappa})$ and $P (\bm{\kappa} \vert \mathbf{L})$ were taken as uniform densities and $P (\mathbf{L})$ was submitted to spatial regularization~\citep{Besag:1986,Scherrer_DIAMOND:2016}, merely reflecting the fact that the brain is a continuous diffusion medium that does not contain actual voxels.

\subsection{In vivo acquisitions} 
\label{Sec_in_vivo}

MRI acquisitions were performed on a clinical 3T system with 45 mT/m maximum gradient amplitude (Ingenia, Philips Healthcare, Best, the Netherlands) using a 32-channel head coil. Imaging was performed on two healthy young male volunteers using a prototype diffusion-weighted spin-echo EPI sequence with numerically optimized~\citep{Sjolund:2015} Maxwell-compensated~\citep{Szczepankiewicz_Maxwell:2019} spherical, planar and linear encoding waveforms. We also attempted to match their respective frequency contents~\citep{Lundell:2019}, but did not sacrifice encoding efficiency to do so. 

Acquisition parameters were: TR=6500 ms, TE=121 ms, spatial resolution=2.5$\times$2.5$\times$2.5 mm$^3$, 48 slices, in-plane acceleration factor=1.9, multiband factor (SENSE)=2, FoV=240$\times$240 mm$^2$, echo spacing=0.82 ms, multi-shell scheme of 45 signal samples at 1$\times b=0$, 6$\times b=0.1$, 6$\times b=0.7$, 12$\times b=1.4$ and 20$\times b=2$ (ms/\textmu$\mathrm{m}^2$) following the scheme suggested in Ref.~\onlinecite{Szczepankiewicz_DIVIDE:2019}, and encoding times of 41 ms (pre), 41 ms (post), 16 ms (pause) for linear gradient waveforms, and 45.33 ms (pre), 41.33 ms (post), 16 ms (pause) for planar and spherical gradient waveforms. All waveforms (including the spherical ones) were rotated \textit{via} a rotation scheme similar to the one used in Ref.~\onlinecite{Szczepankiewicz_data:2019}. The acquired images were corrected for motion and Eddy currents by applying an in-house topup-eddy procedure based on a common averaged $b=0$ image for the linear, planar and spherical parts of the dataset independently. They were then resampled to 2$\times$2$\times$2 mm$^3$ using linear interpolation. Other topup-eddy strategies for tensor-valued diffusion encoded data can be found in the literature \citep{Nilsson:2015}. The signal-to-noise ratio (SNR) of the \textit{in vivo} dataset was estimated to around 40 in the corona radiata using the method described in the Supplemental Material of Ref.~\onlinecite{Szczepankiewicz_DIVIDE:2019} on our $b=0.7$~ms/\textmu$\mathrm{m}^2$ spherical data. Finally, in order to evaluate the impact of introducing tensor-valued diffusion encoding within Magic DIAMOND, the acquired volumes were combined so as to create the three following encoding combinations: 
\begin{itemize}
\item[$\bullet$] LL: 45 linear $+$ 45 linear b-tensors,
\item[$\bullet$] LP: 45 linear $+$ 45 planar b-tensors,
\item[$\bullet$] LS: 45 linear $+$ 45 spherical b-tensors.
\end{itemize}

\subsection{Numerical simulations} 
\label{Sec_in_silico}

\begin{figure}[h!]
\begin{center}
\includegraphics[width=20pc]{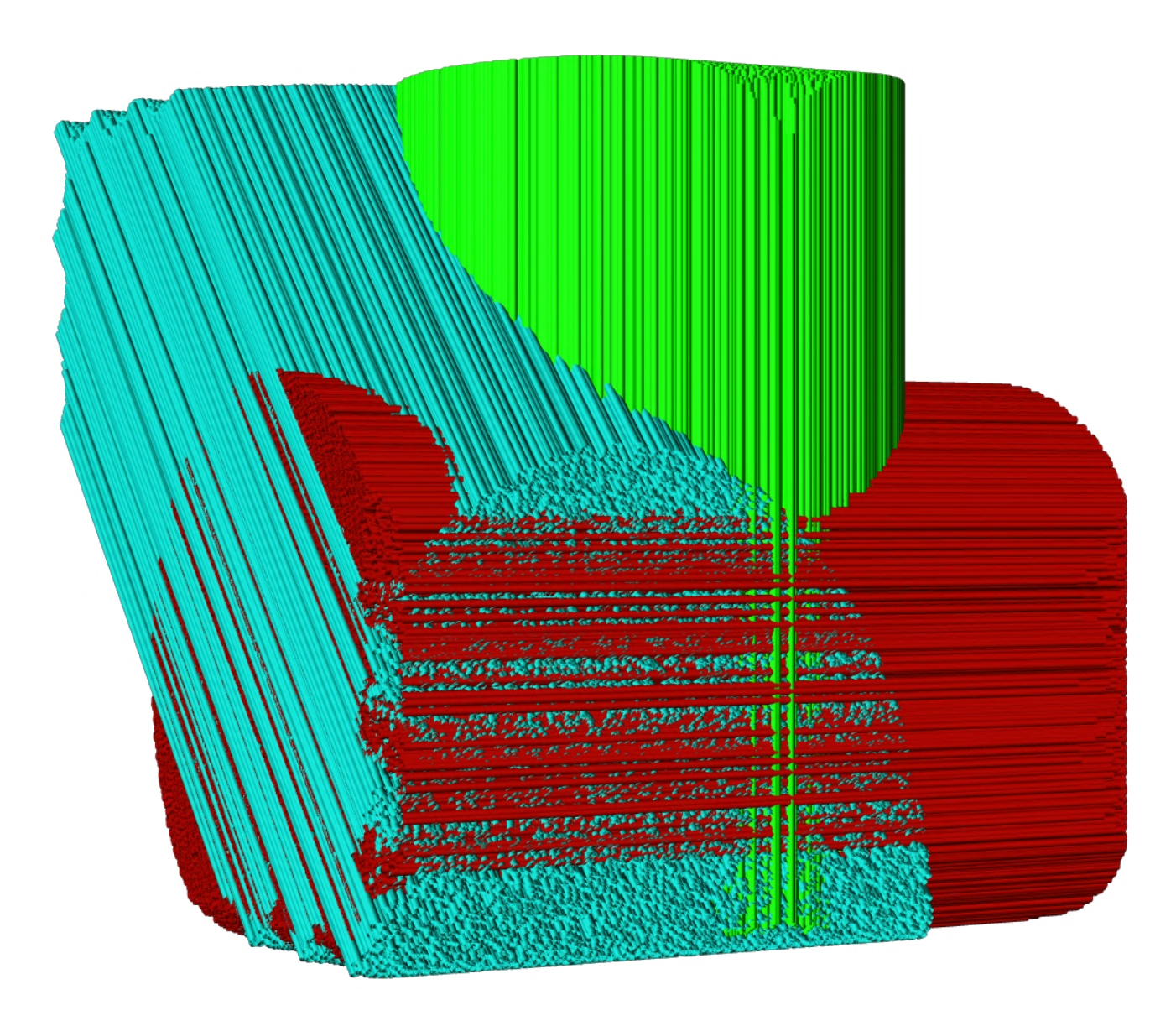}
\caption{Illustration of the Fiberfox numerical phantom described in section~\ref{Sec_in_silico} and used to evaluate Magic DIAMOND in section~\ref{Sec_results_in_silico}. The phantom is made out of three fascicles: one along the $x$-axis (red), one along the $y$-axis (green), and one crossing with them at a 45$^\circ$ angle between the $y$- and $z$- axes (blue). Each fascicle is composed of 900 000 fibers characterized by the axial and radial diffusivities $D_\parallel = $~1.7~{\textmu}m$^2$/ms and $D_\perp = $~0.4~{\textmu}m$^2$/ms, respectively. The radial distribution of these fibers follows a normal distribution and the remaining extra-fiber space is filled up with free water of isotropic diffusivity $D_\mathrm{FW} =$~3~{\textmu}m$^2$/ms.}
\label{Figure_phantom}
\end{center}
\end{figure}

Numerical simulations were performed using a modified version of Fiberfox~\citep{Neher:2014} that allows for tensor-valued diffusion encoding. We generated a medium composed of three partially crossing fascicles surrounded by water, described by the free isotropic diffusivity $D_\mathrm{FW} = 3$~\textmu$\mathrm{m}^2/\mathrm{ms}$. Two fascicles span along the $x$- and $y$- axes, respectively, and the third fascicle crosses with them at a $45^\circ$ angle between the $y$- and $z$- axes. Each fascicle is composed of $900\,000$ fibers characterized by the axial and radial diffusivities $D_\parallel = 1.7$~\textmu$\mathrm{m}^2/\mathrm{ms}$ and $D_\perp = 0.4$~\textmu$\mathrm{m}^2/\mathrm{ms}$, respectively \citep{Pierpaoli:1996, Descoteaux:2006}. The radial distribution of these fibers follows a normal distribution and the remaining extra-fiber space is filled up with free water. Our Fiberfox phantom is illustrated in Fig.~\ref{Figure_phantom}. 

Datasets were simulated for linear, planar and spherical diffusion encodings at two SNRs, namely the $\mathrm{SNR}=40$ of our \textit{in vivo} dataset and the ideal infinite SNR. The Rician noise of the $\mathrm{SNR}=40$ dataset was generated as follows:
\begin{equation}
\mathcal{S}_\mathrm{noisy}(\mathbf{b}) = \mathcal{S}_{0,\mathrm{true}} \sqrt{\left( \frac{\mathcal{S}_\mathrm{true}(\mathbf{b})}{\mathcal{S}_{0,\mathrm{true}}} + \frac{\nu}{\mathrm{SNR}} \right)^2 + \left( \frac{\nu^\prime}{\mathrm{SNR}} \right)^2}\, ,
\label{Eq_Rician_noise}
\end{equation}
where $\mathcal{S}_{\mathrm{true}}$ and $\mathcal{S}_{0,\mathrm{true}}=\mathcal{S}_{\mathrm{true}}(\mathbf{b}=\mathbf{0})$ come from the infinite-SNR (noise-free) dataset, and $\nu$ and $\nu^\prime$ are drawn from a normal distribution with zero mean and unit standard deviation. We set the same acquisition parameters and considered the same encoding combinations as those described in section~\ref{Sec_in_vivo}. To assess whether or not two encoding combinations yield distinct estimated values for a given metric, we used non-parametric Mann-Whitney U-tests \citep{McKnight_Najab:2010}, equivalent to two-sided Wilcoxon rank sum tests.

\subsection{Precision on parameter estimation with stratified bootstrap} 
\label{Sec_bootstrap}

The evaluation of diffusion models with \textit{in vivo} data is challenging because the underlying ground truth is not known. We compared the various encoding strategies by evaluating their precision on parameter estimation. Intuitively, the precision is high when multiple acquisitions of the same object leads to similar results and does not depend on noise realizations. Assessing estimation uncertainty allowed us to test whether the additional information provided by the tensor-valued diffusion encoding better constrains the estimation problem, while keeping a constant number of DW images. 

The typical way to compute estimation uncertainty is to repeat the measurement of the same object, estimate parameters for each and assess the variance in parameters. This would unfortunately lead to unrealistic scan time in DW-MRI. We instead used \textit{stratified bootstrap}, which amounts to 1) acquire two repetitions of the same DW-MRI experiment, and 2) create a large number of "virtual" acquisitions by randomly choosing, for each diffusion gradient direction, one of the two repetitions. Unlike wild bootstrap~\citep{Whitcher:2008} and repetition bootstrap~\citep{Chung:2006}, stratified bootstrap allows for the creation of a large number of noise realizations without any explicit noise modeling.
Specifically, we acquired twice each of the datasets described in section~\ref{Sec_in_vivo}, namely four sets of 45 linear b-tensors (L1, L2, L3, L4), two sets of 45 planar b-tensors (P1, P2), and two sets of 45 spherical b-tensors (S1, S2), giving the following combinations: 
\begin{itemize}
\item[$\bullet$] LL: [L1 $\longleftrightarrow$ L2] $+$ [L3 $\longleftrightarrow$ L4],
\item[$\bullet$] LP: [L1 $\longleftrightarrow$ L2] $+$ [P1 $\longleftrightarrow$ P2],
\item[$\bullet$] LS: [L1 $\longleftrightarrow$ L2] $+$ [S1 $\longleftrightarrow$ S2],
\end{itemize}
where $\longleftrightarrow$ denotes random choice between two datasets on each acquired direction and $+$ denotes dataset concatenation. We generated 100 realizations of each of these combinations. Medians and interquartile ranges (IQRs) of estimated parameters across these realizations were then computed within a white-matter mask to quantify their average value and variance, respectively. We compared the results from only linear b-tensors (LL, equivalent to the original DIAMOND formulation) to those arising from combinations of linear, planar and spherical b-tensors (LP and LS).

\subsection{Visualization of the optimization landscape with in vivo data}
\label{Sec_cost_function}

The estimation of a DCI model is known to be a challenging inversion problem involving multidimensional functionals with numerous local minima.
We evaluated the optimization landscape by investigating the shape of the cost function used for parameter estimation~\citep{Scherrer:2012} in the parameter space. This investigation was carried out in two voxels of interest, a typical one-fiber voxel in the corpus callosum and a typical three-fiber voxel in the centrum semiovale, after evaluating Magic DIAMOND on the following sets of 180 b-tensors (covering all our acquired signals, see section~\ref{Sec_bootstrap}):
\begin{itemize}
\item[$\bullet$] LL: L1 $+$ L2 $+$ L3 $+$ L4,
\item[$\bullet$] LP: L1 $+$ L2 $+$ P1 $+$ P2,
\item[$\bullet$] LS: L1 $+$ L2 $+$ S1 $+$ S2,
\end{itemize}
where $+$ denotes dataset concatenation. Finally, we compared the different cost function shapes extracted from the LL, LP and LS acquisition schemes.

\subsection{Tractography and fascicle-metric mapping}
\label{Sec_tractography}
While computing whole-brain maps is the most common approach with single-fascicle models, the lack of one-to-one correspondence between neighbouring voxels with multi-fascicle models makes them more challenging to evaluate. For instance, there may be fascicles that are not present in all voxels (one-to-zero correspondence) or fascicles that are represented by different number of compartments in different voxels (one-to-many correspondence). We therefore evaluated the impact of tensor-valued diffusion encoding using a fixel-based (fascicle-element based) analysis \citep{Raffelt:2015}, by delineating tract streamlines with tractography and by extracting quantities related to the most aligned fascicle compartment along the streamlines.

Multi-peak tractography \citep{Chamberland:2014} was performed on the fFA-weighted peaks extracted from Magic-DIAMOND evaluation on the datasets defined in section~\ref{Sec_cost_function}. To obtain dense tractograms, we upsampled our peaks to 1$\times$1$\times$1 mm$^3$ using nearest-neighbor interpolation, used ten seeds per voxel 
and generated an ensemble tractogram \citep{Takemura:2016} for each dataset with three separate tractography runs with acceptance angles of $45^\circ$, $60^\circ$ and $75^\circ$. The output tracks were finally filtered using the recent "Recobundles" algorithm~\citep{Garyfallidis:2018} that relies on shape analysis and anatomical priors to recognize and extract bundles of interest, namely the corpus callosum, the arcuate fasciculus and the corticospinal tract, three tracts that intersect in the centrum semiovale, making them difficult to reconstruct in their full extent \citep{Rheault:2019}. Each bundle tractogram was then used as a skeleton onto which metrics associated to the Magic-DIAMOND fascicle most aligned with any given local track orientation (sub-voxel tract segment) can be color-mapped. Performing this process across all bootstrap realizations described in section~\ref{Sec_bootstrap} enabled color-mapping of the median and interquartile range of any metric \textit{along} streamlines. 

\section{Results}
\label{Sec_Results}

\subsection{In silico data}
\label{Sec_results_in_silico}

\begin{figure}[ht!]
\begin{center}
\includegraphics[width=20pc]{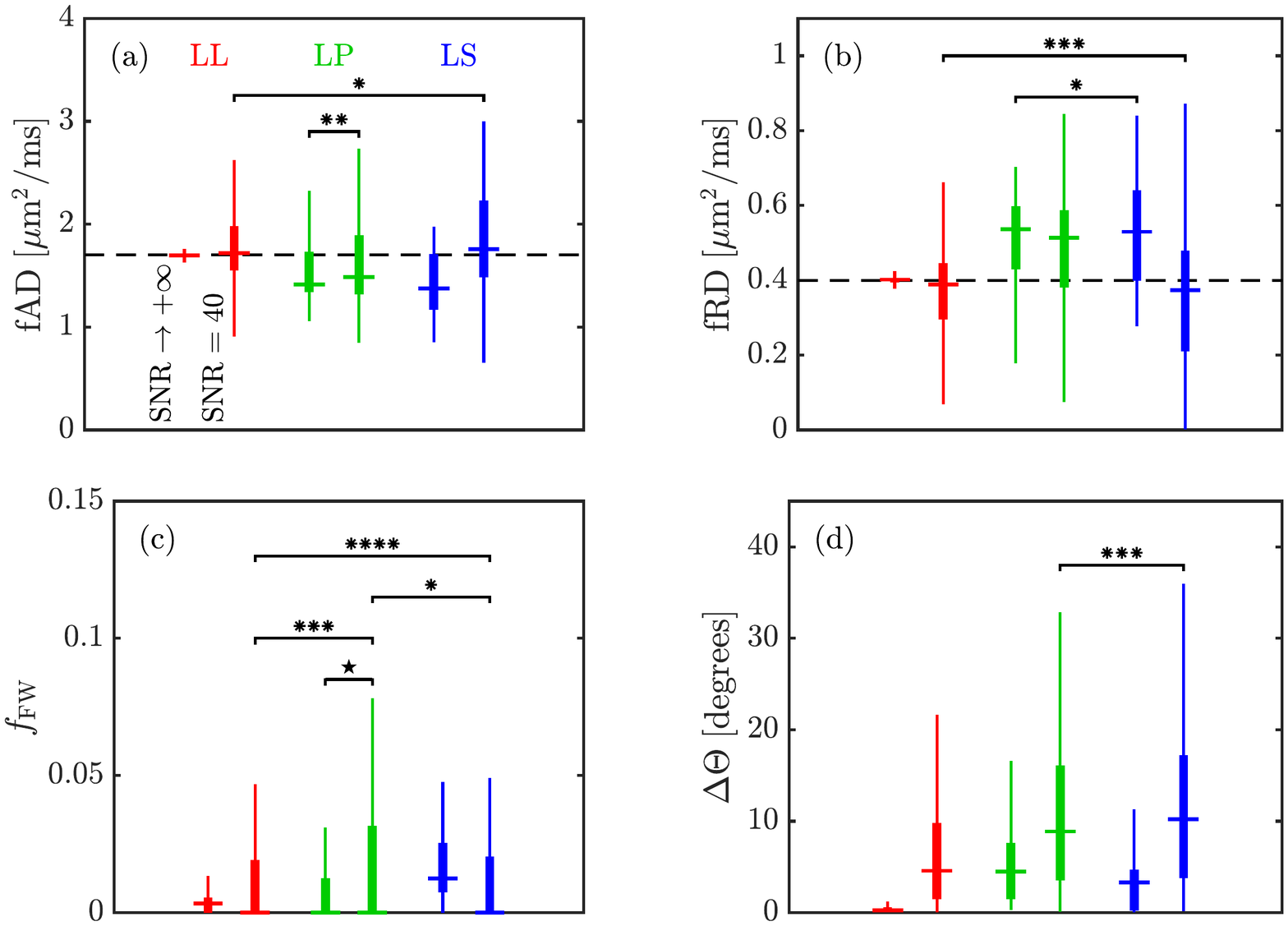}
\caption{In silico evaluation of Magic DIAMOND within a mask corresponding to the fascicle aligned with $x$ in the Fiberfox phantom illustrated in Fig.~\ref{Figure_phantom} for LL (red), LP (green) and LS (blue) at infinite SNR and SNR=40. The estimated metrics are the fascicle axial diffusivity (fAD, panel (a)), the fascicle radial diffusivity (fRD, panel (b)), the FW signal fraction $f_\mathrm{FW}$ (panel (c)), and the angular deviation $\Delta \Theta$ (panel (d)). $\Delta \Theta$ is defined voxel-wise as the average of the angles between each Magic-DIAMOND estimated fascicle and its closest Fiberfox fiber bundle. Results of non-parametric Mann-Whitney U-tests applied between metric distributions are reported in order to assess whether or not two Magic-DIAMOND estimations yield statistically distinct metrics. Such tests consider the null hypothesis $\mathcal{H}_0$ that data in two Magic-DIAMOND evaluations are sampled from identically shaped non-median-shifted continuous distributions, against the alternative that they are not. The $p$-values resulting from these tests inform on the acceptance or rejection of $\mathcal{H}_0$ at a certain significance level: $*\equiv 0.01\leq p < 0.05$, $**\equiv 0.05\leq p < 0.1$, $***\equiv 0.1\leq p < 0.15$, $****\equiv 0.15 \leq p < 0.2$, and $\star \equiv 0.6 \leq p \leq 0.7$.}
\label{Figure_Fiberfox}
\end{center}
\end{figure}

Fig.~\ref{Figure_Fiberfox} compares physical quantities generated within a mask corresponding to the fascicle aligned with $x$ in our Fiberfox phantom (see Fig.~\ref{Figure_phantom}) with their Magic-DIAMOND estimations for the encoding combinations LL, LP and LS at infinite SNR and at the finite $\mathrm{SNR}=40$ characterizing the \textit{in vivo} dataset described in section~\ref{Sec_in_vivo}. These physical quantities comprise the fascicle axial diffusivity (fAD), the fascicle radial diffusivity (fRD), the FW signal fraction $f_\mathrm{FW}$, and the angular deviation $\Delta \Theta$. This last quantity is defined voxel-wise as the average of the angles between each Magic-DIAMOND estimated fascicle and its closest Fiberfox fiber bundle.
	
\subsection{Precision on parameter estimation by stratified bootstrap on in vivo data}

\begin{figure}[ht!]
\begin{center}
\includegraphics[width=20pc]{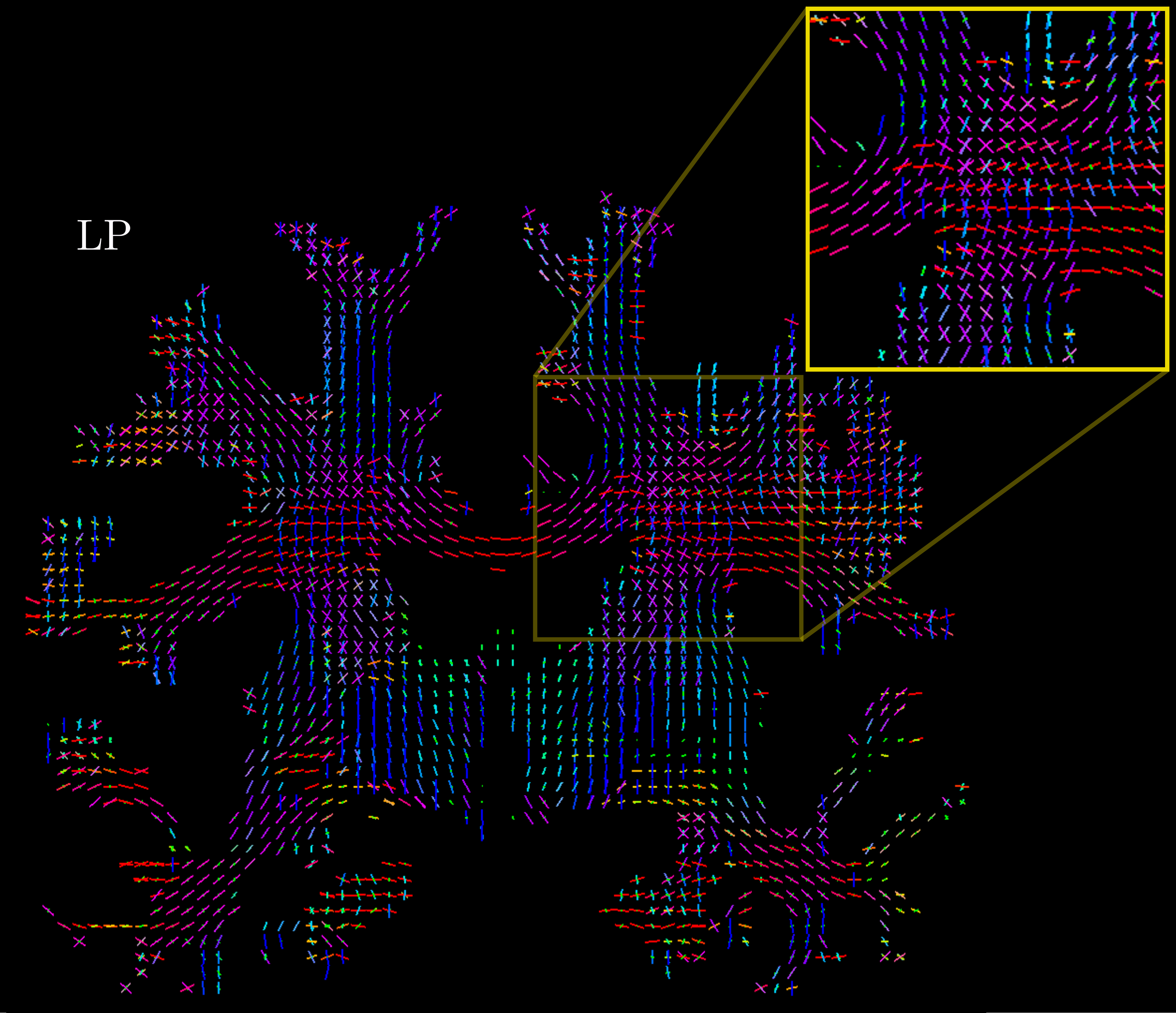}
\caption{Local orientations of the fascicles estimated by Magic DIAMOND for our LP \textit{in vivo} dataset in a white-matter masked coronal slice. These orientations, consistent with the expected anatomy, were computed after one stratified bootstrap signal realization. \textbf{Inset:} Zoom on the three-way crossing of the centrum semiovale.}
\label{Figure_local}
\end{center}
\end{figure}

\begin{figure}[ht!]
\begin{center}
\includegraphics[width=20pc]{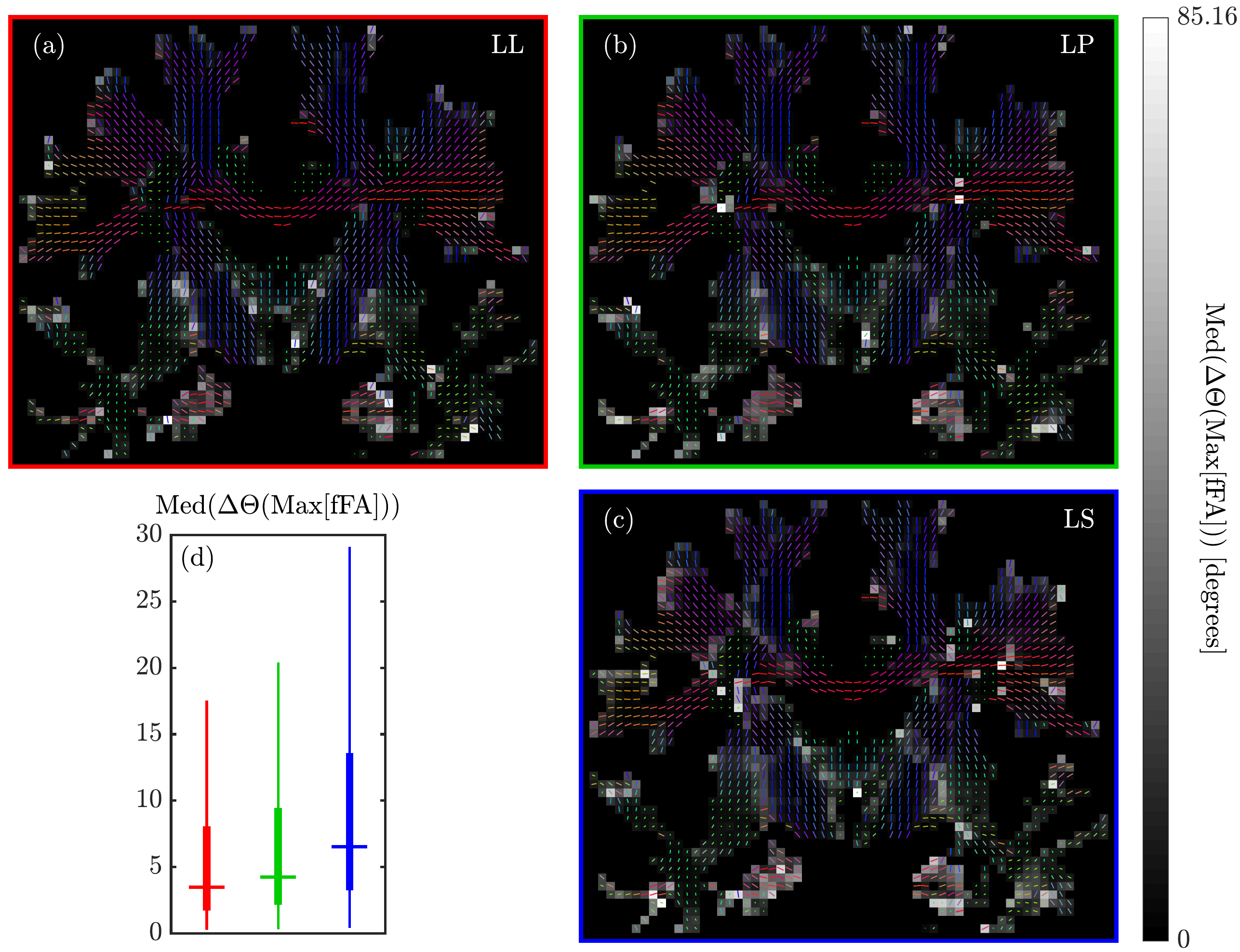}
\caption{Median angular deviation $\mathrm{Med}(\Delta\Theta(\mathrm{Max[fFA]}))$ (in degrees) across 100 stratified bootstrap runs of Magic DIAMOND, computed for the orientation of the intra-voxel fascicle with maximal fFA. For a given stratified bootstrap run, $\Delta\Theta(\mathrm{Max[fFA]})$ was computed voxel-wise as the angular difference between the orientation of the Max[fFA] fascicle within this run and the median orientation of this fascicle across stratified bootstrap runs. $\mathrm{Med}(\Delta\Theta(\mathrm{Max[fFA]}))$ is here presented in two ways. Panels (a), (b) and (c): White-matter masked coronal views of $\mathrm{Med}(\Delta\Theta(\mathrm{Max[fFA]}))$ (greyscale maps) for LL, LP and LS, respectively, with superimposed median orientations of the Max[fFA] fascicles across stratified bootstrap runs (orientation-colored peaks). Panel (d): Distribution of the values of $\mathrm{Med}(\Delta\Theta(\mathrm{Max[fFA]}))$ within our whole-brain white-matter mask for LL (red), LP (green) and LS (blue). The main features of these distributions, \textit{i.e.} medians and interquartile ranges (IQRs), are quantified as boxplots.}
\label{Figure_orientations}
\end{center}
\end{figure}

A typical map of local orientations in a coronal slice is presented in Fig.~\ref{Figure_local}, featuring orientations that are consistent with the known anatomy, from the one-fiber corpus callosum to the three-way crossing of the centrum semiovale. To assess the uncertainty on estimation for these orientations within white matter, Fig.~\ref{Figure_orientations} presents the distribution of the median angular deviation $\mathrm{Med}(\Delta\Theta(\mathrm{Max[fFA]}))$ for the orientation of the intra-voxel fascicle with maximal fFA, chosen as a proxy to evaluate the same fascicle orientation, across 100 stratified bootstrap runs of Magic DIAMOND for LL, LP and LS.

\begin{figure}[ht!]
\begin{center}
\includegraphics[width=20pc]{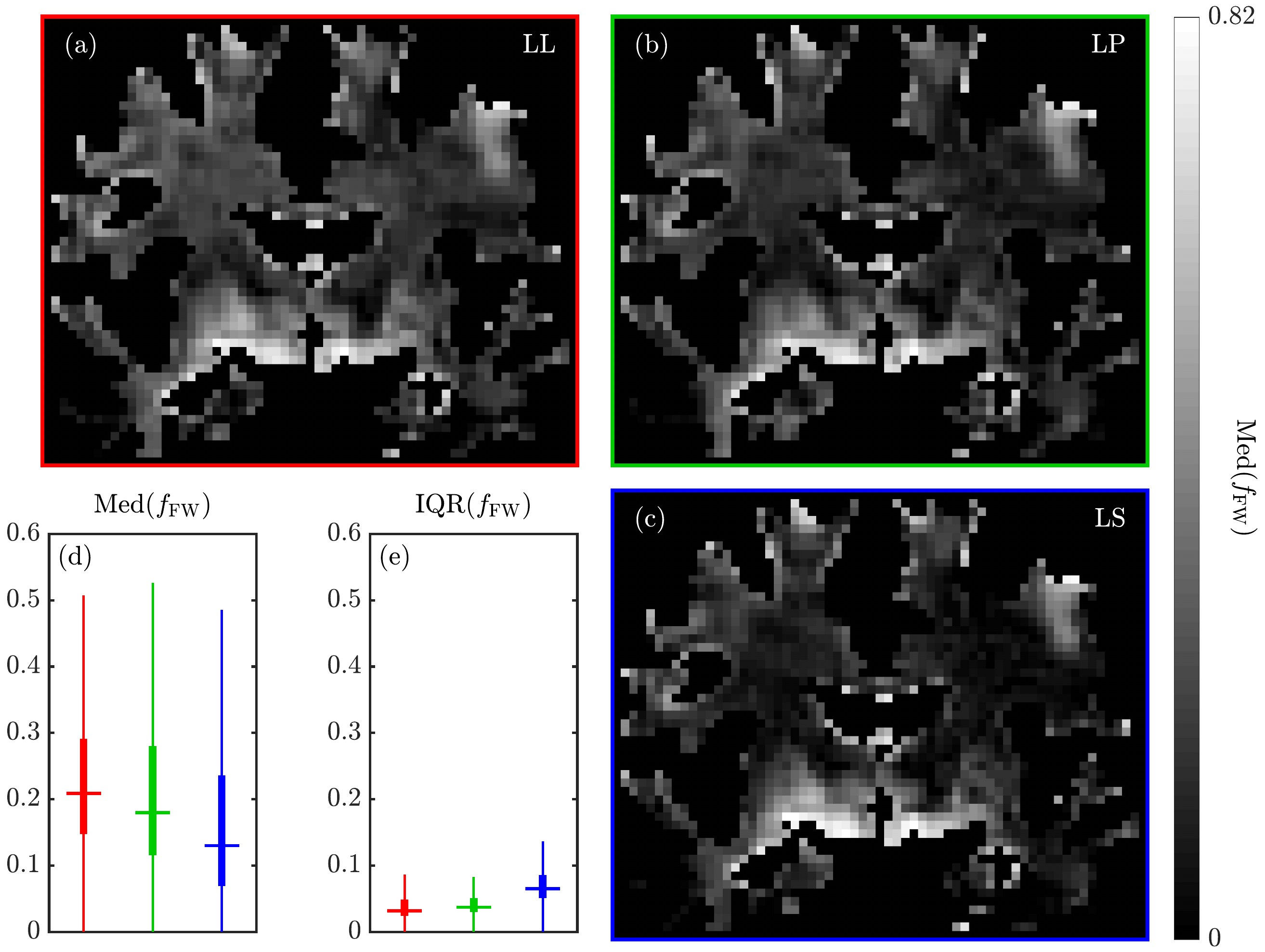}
\caption{Median value (Med) and interquartile range (IQR) of the free water signal fraction $f_\mathrm{FW}$ over 100 stratified bootstrap runs of Magic DIAMOND. Panels (a), (b) and (c): White-matter masked coronal views of $\mathrm{Med}(f_\mathrm{FW})$ (greyscale maps) for LL, LP and LS, respectively. Panels (d) and (e): Distributions of the values of $\mathrm{Med}(f_\mathrm{FW})$ and $\mathrm{IQR}(f_\mathrm{FW})$, respectively, within our whole-brain white-matter mask for LL (red), LP (green) and LS (blue). The main features of these distributions, \textit{i.e.} medians and interquartile ranges, are quantified as boxplots.}
\label{Figure_fFW}
\end{center}
\end{figure}

\begin{figure}[ht!]
\begin{center}
\includegraphics[width=20pc]{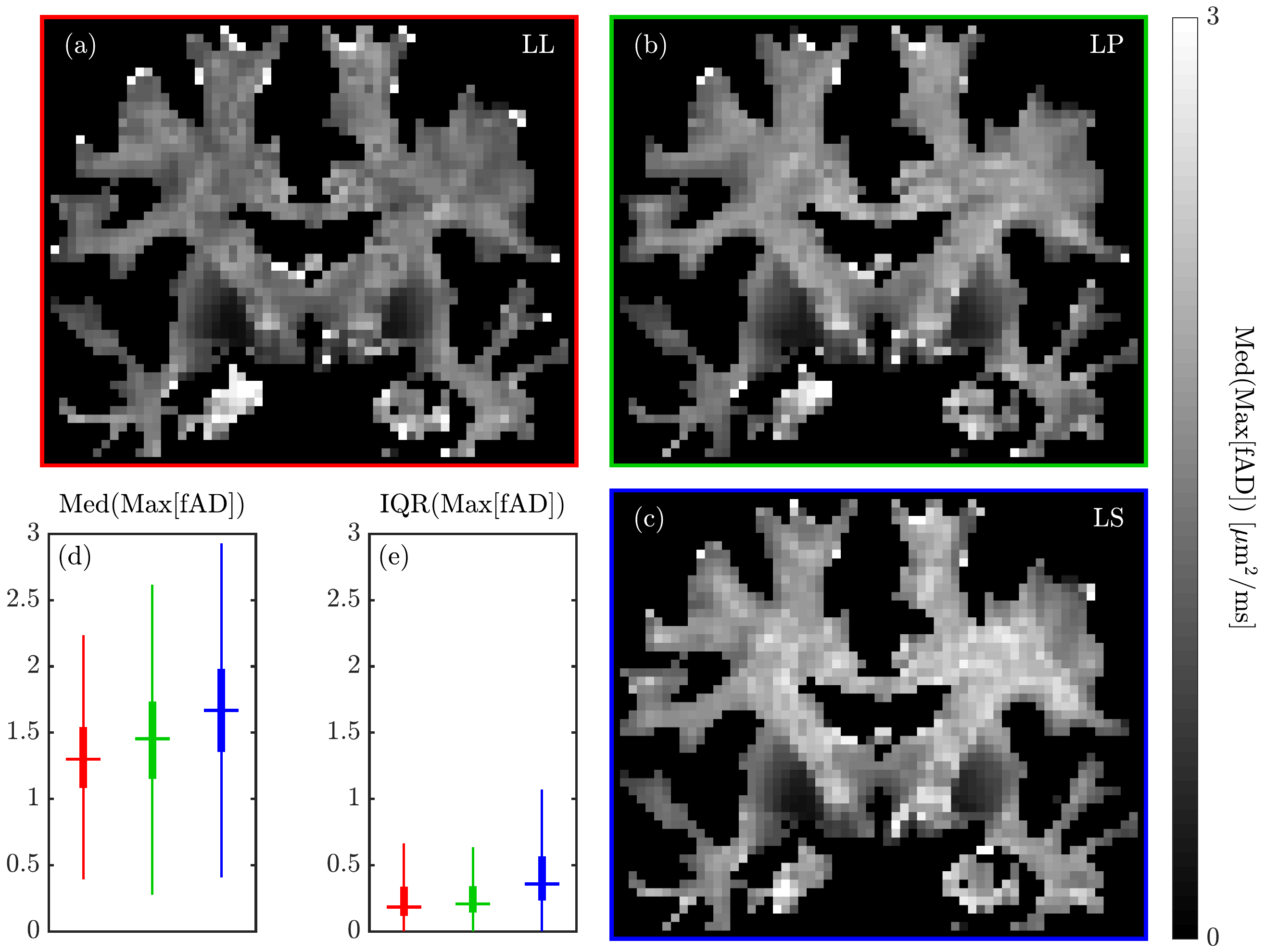}
\caption{Median value (Med) and interquartile range (IQR) of the maximal fascicle axial diffusivity ($\mathrm{Max[fAD]}$, in {\textmu}m$^2$/ms) over 100 stratified bootstrap runs of Magic DIAMOND. Layout conventions are identical to those of Fig.~\ref{Figure_fFW}.}
\label{Figure_fAD}
\end{center}
\end{figure}

\begin{figure}[ht!]
\begin{center}
\includegraphics[width=20pc]{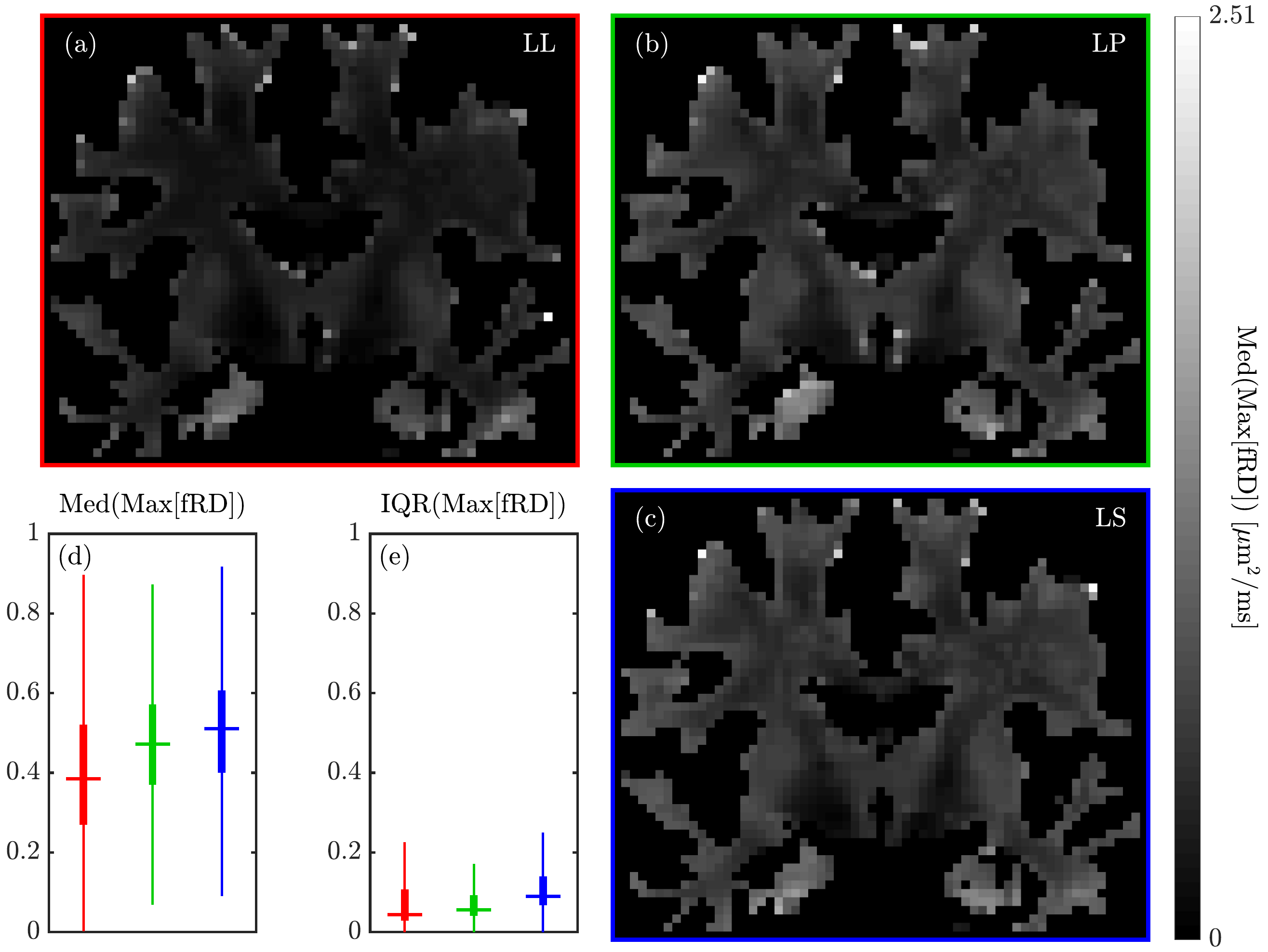}
\caption{Median value (Med) and interquartile range (IQR) of the maximal fascicle radial diffusivity ($\mathrm{Max[fRD]}$, in {\textmu}m$^2$/ms) over 100 stratified bootstrap runs of Magic DIAMOND. Layout conventions are identical to those of Fig.~\ref{Figure_fFW}.}
\label{Figure_fRD}
\end{center}
\end{figure}

Figs.~\ref{Figure_fFW}-\ref{Figure_fAD}-\ref{Figure_fRD} report on the white-matter distributions of three scalar metrics estimated throughout the stratified bootstrap experiment: the FW signal fraction $f_\mathrm{FW}$, the maximal fascicle axial diffusivity Max[fAD] and the maximal fascicle radial diffusivity Max[fRD], respectively. The Max[$\,\cdot\,$] operator was used for fAD and fRD as a proxy to identify between each bootstrap realization the same fascicle and to show a single whole-brain map of Max[fAD] and Max[fRD] in Figs.~\ref{Figure_fAD}-\ref{Figure_fRD}. 

\subsection{Fascicle metric mapping on tractograms}

\begin{figure*}[ht!]
\begin{center}
\includegraphics[width=42pc]{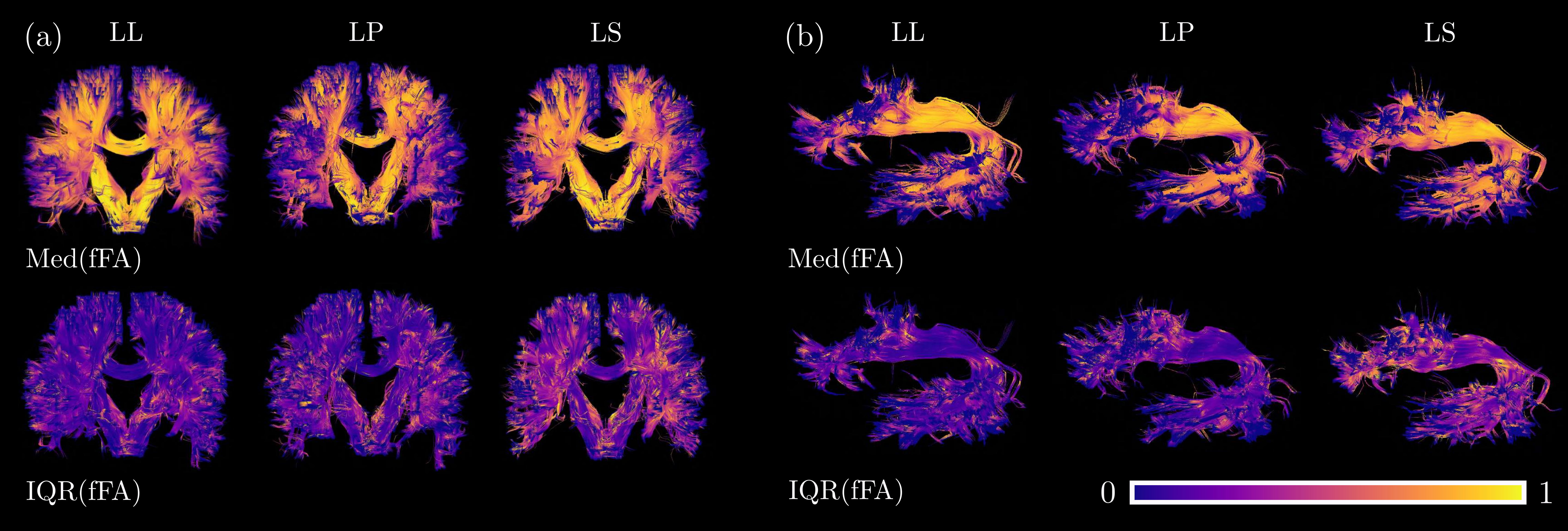}
\caption{Color-mapping of Magic DIAMOND metrics onto bundles of interest. Panel (a): Coronal view of the tractograms associated to the corpus callosum, the arcuate fasciculus and the corticospinal tract. Panel (b): Sagittal view of the tractograms associated to the left arcuate fasciculus. The median (Med) and interquartile range (IQR) across all stratified bootstrap realizations of the fFA associated to the Magic-DIAMOND fascicle most aligned with the local tract orientation were color-mapped along these bundles for each encoding combination described in sections~\ref{Sec_cost_function}-\ref{Sec_tractography}.}
\label{Figure_tracto}
\end{center}
\end{figure*}

Fig.~\ref{Figure_tracto} shows tractograms where we color-mapped the median and interquartile range of fFA across all bootstrap realizations along the corpus callosum, the arcuate fasciculus and the corticospinal tract. 

\subsection{Visualization of the optimization landscape with in vivo data}

\begin{figure*}[ht!]
\begin{center}
\includegraphics[width=42pc]{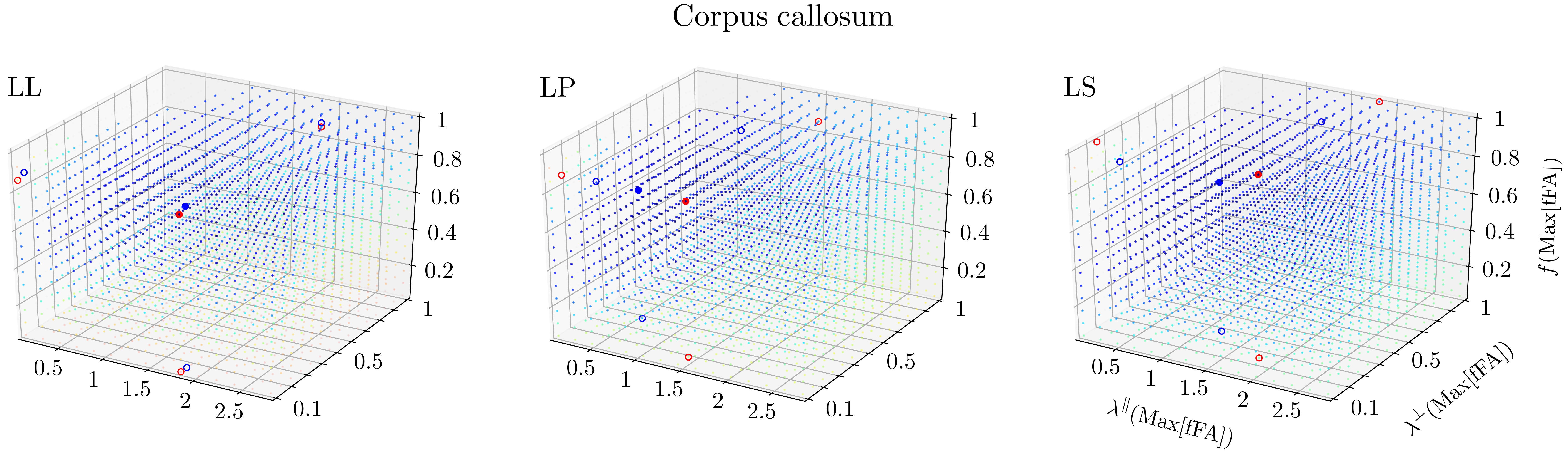}
\caption{Visualization of the minimization landscape in a corpus-callosum voxel for the LL, LP and LS signal combinations described in section~\ref{Sec_cost_function}. The cost function was computed on sampled grid points in a three-dimensional parameter subspace encapsulating the axial diffusivity $\lambda^\parallel$, the radial diffusivity $\lambda^\perp$ and the signal fraction $f$ of the single fascicle found in this corpus-callosum voxel (which has maximal fFA by default). Color encodes the value of the cost function, from its minimal value (blue) to its maximal value (red). Opacity decreases with increasing cost function to focus on regions of the parameter subspace that are relevant for minimization. Diffusivities are expressed in~{\textmu}m$^2$/ms. While the filled blue point is associated to the minimal cost function found by Magic DIAMOND, the filled red point corresponds to the global minimum in the sampled parameter subspace. Empty circles denote projections of the filled points.}
\label{Figure_cost_function_CC_1}
\end{center}
\end{figure*}

\begin{figure*}[ht!]
\begin{center}
\includegraphics[width=42pc]{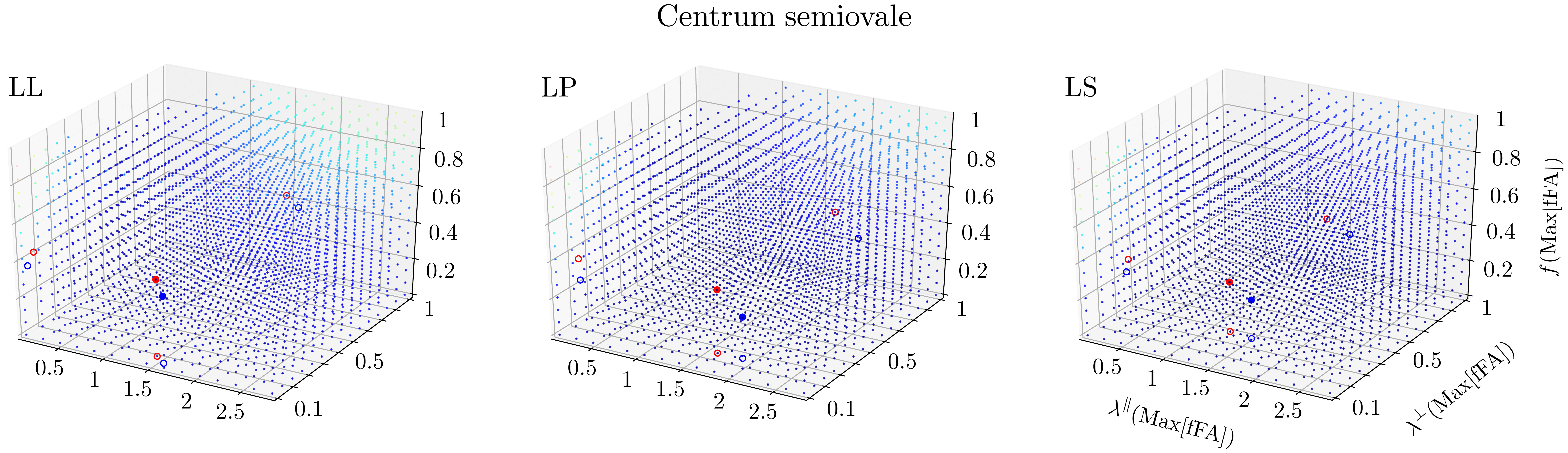}
\caption{Visualization of the minimization landscape in a centrum-semiovale voxel for the LL, LP and LS signal combinations described in section~\ref{Sec_cost_function}. The cost function was computed on sampled grid points in a three-dimensional parameter subspace encapsulating the axial diffusivity $\lambda^\parallel$, the radial diffusivity $\lambda^\perp$ and the signal fraction $f$ of the fascicle with maximal fFA in this centrum-semiovale voxel that contained three intra-voxel fascicles. Layout conventions are identical to those of Fig.~\ref{Figure_cost_function_CC_1}.}
\label{Figure_cost_function_CS_1}
\end{center}
\end{figure*}

\begin{figure*}[ht!]
\begin{center}
\includegraphics[width=42pc]{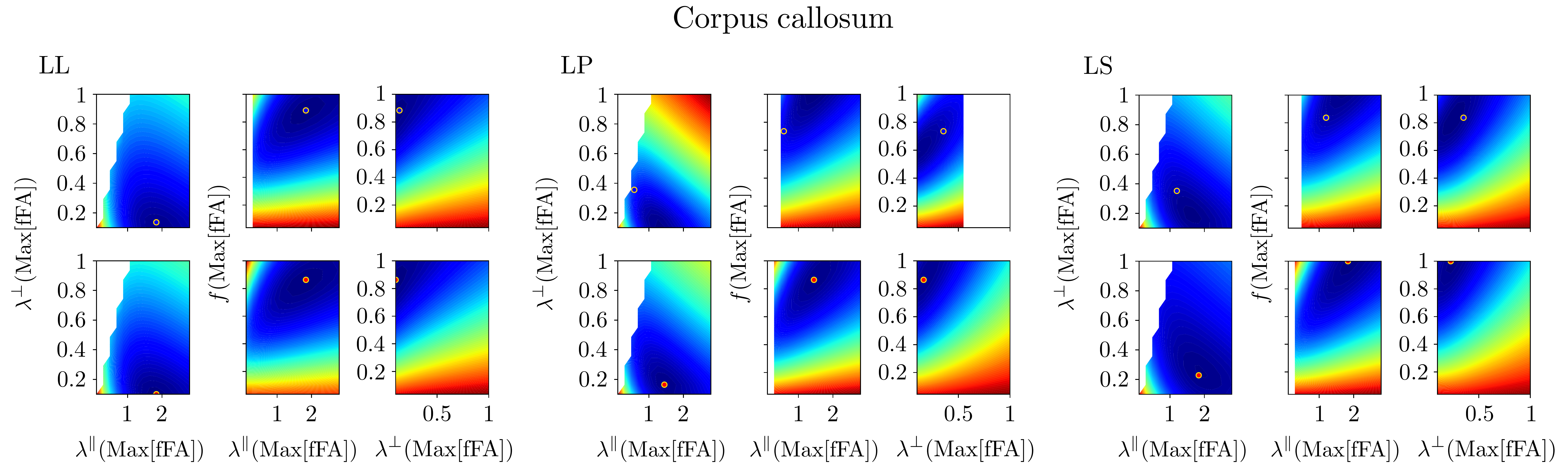}
\caption{Visualization of two-dimensional cuts of interest in the three-dimensional minimization landscape shown in Fig.~\ref{Figure_cost_function_CC_1} for the LL, LP and LS signal combinations described in section~\ref{Sec_cost_function}. The color maps indicate the value of the cost function, from its minimal value (blue) to its maximal value (red), within two-dimensional projections intersecting the Magic DIAMOND minimum (blue point, top) and the global minimum in the aforementioned three-dimensional subspace (red point, bottom). White areas correspond to unphysical fiber configurations where $\lambda^\perp > \lambda^\parallel$. Diffusivities are expressed in~{\textmu}m$^2$/ms.}
\label{Figure_cost_function_CC_2}
\end{center}
\end{figure*}

\begin{figure*}[ht!]
\begin{center}
\includegraphics[width=42pc]{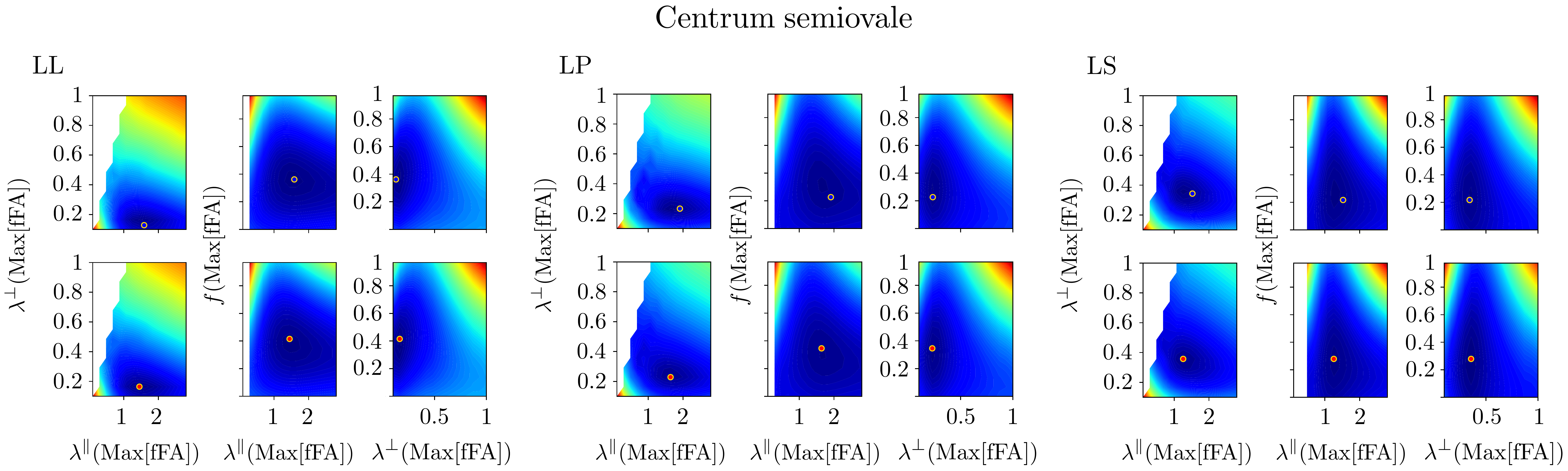}
\caption{Visualization of two-dimensional cuts of interest in the three-dimensional minimization landscape shown in Fig.~\ref{Figure_cost_function_CS_1} for the LL, LP and LS signal combinations described in section~\ref{Sec_cost_function}. Layout conventions are identical to those of Fig.~\ref{Figure_cost_function_CC_2}.}
\label{Figure_cost_function_CS_2}
\end{center}
\end{figure*}

Figs.~\ref{Figure_cost_function_CC_1}-\ref{Figure_cost_function_CS_1}-\ref{Figure_cost_function_CC_2}-\ref{Figure_cost_function_CS_2} provide visualizations of the minimization space with LL, LP and LS. Specifically, they show the three-dimensional subspace of parameters covering various axial diffusivities $\lambda^\parallel$, radial diffusivities $\lambda^\perp$ and signal fractions $f$ for the intra-voxel Max[fFA] fascicle (Figs.~\ref{Figure_cost_function_CC_1}-\ref{Figure_cost_function_CS_1}), and two-dimensional cuts of this subspace (Figs.~\ref{Figure_cost_function_CC_2}-\ref{Figure_cost_function_CS_2}), in two different voxels of interest: one in the corpus callosum (CC, single fiber), the other in the centrum semiovale (CS, three-way crossing). Note that the Magic-DIAMOND minimization was in fact performed in a six-dimensional space for the corpus callosum and in a sixteen-dimensional space for the centrum semiovale (one parameter for the free-water compartment, five parameters for each of the three fascicles). 

\section{Discussion}
\label{Sec_Discussion}

The \textit{in silico} results presented at finite SNR=40 and infinite SNR in Fig.~\ref{Figure_Fiberfox} report on the accuracy and precision of Magic DIAMOND's estimations for various encoding combinations (LL, LP and LS, described in section~\ref{Sec_in_vivo}) within a numerical phantom that matches DIAMOND's compartmental assumptions (see section~\ref{Sec_in_silico} and Fig.~\ref{Figure_phantom}). The LL encoding combination yields the most accurate and precise Magic-DIAMOND estimations, both at infinite and finite SNRs. At constant SNR, precision decreases when going from LL to LP, and from LP to LS. In terms of accuracy, LP and LS present infinite-SNR biases that are absent in LL. While LS's accuracy improves at finite SNR when estimating non-orientational information, highlighting an acute effect of noise in the data, LP's accuracy exhibits a resilience to noise comparable to that of LL's. Our non-parametric Mann-Whitney U-tests indicate that there is no statistical difference between the estimations yielded by LL and LS, in contrast with those yielded by LP for most metrics.

\textit{In vivo}, uncertainty on parameter estimation is evaluated for Magic DIAMOND \textit{via} the stratified bootstrap procedure detailed in section~\ref{Sec_bootstrap}. From the standpoint of diffusion orientation information alone, while Fig.~\ref{Figure_local} features estimated local orientations that are consistent with the known anatomy, Fig.~\ref{Figure_orientations} shows that angular uncertainty increases when going from LL to LP, and from LP to LS, in agreement with the \textit{in silico} results of Fig.~\ref{Figure_Fiberfox}. Spatially, angular uncertainty tends to maximize in expected areas, \textit{i.e.} cortical grey matter, deep grey nuclei, and crossing-fascicle regions. 

From the standpoint of non-orientational metrics, Figs.~\ref{Figure_fFW}-\ref{Figure_fAD}-\ref{Figure_fRD} demonstrate that while LP presents similar uncertainties on estimation of these parameters compared to LL, as quantified by the median values of the white-matter distributions of $\mathrm{IQR}(f_\mathrm{FW})$, $\mathrm{IQR}(\mathrm{Max[fAD]})$ and $\mathrm{IQR}(\mathrm{Max[fRD]})$ (panels (e) of Figs.~\ref{Figure_fFW}-\ref{Figure_fAD}-\ref{Figure_fRD}), LS exhibits a rather substantial increase in these uncertainties. The median values of the white-matter distributions of $\mathrm{Med}(f_\mathrm{FW})$, $\mathrm{Med}(\mathrm{Max[fAD]})$ and $\mathrm{Med}(\mathrm{Max[fRD]})$ (panels (d) of Figs.~\ref{Figure_fFW}-\ref{Figure_fAD}-\ref{Figure_fRD}) underline the overall changes in Magic DIAMOND's estimations between diffusion-encoding strategies. Both LP and LS tend to yield lower values of $f_\mathrm{FW}$ and higher values of Max[fAD] and Max[fRD] than LL. Although not shown here, this increase in Max[fAD] and Max[fRD] leads to an increase in maximal fascicle mean diffusivity (Max(fMD)), from $\mathrm{Med(Max[fMD])}\simeq 0.65$~\textmu$\mathrm{m}^2/\mathrm{ms}$ (LL) to $\mathrm{Med(Max[fMD])}\simeq 0.8$~\textmu$\mathrm{m}^2/\mathrm{ms}$ (LS), but rather constant maximal fascicle fractional anisotropy $0.65\leq\mathrm{Med(Max[fFA])}\leq 0.7$. 
We acknowledge that a few discrepancies exist between the \textit{in silico} results of Fig.~\ref{Figure_Fiberfox} and the \textit{in vivo} results of Figs.~\ref{Figure_fFW}-\ref{Figure_fAD}-\ref{Figure_fRD}, which may originate from an insufficient match between our simulated phantom and the microstructural complexity of actual \textit{in vivo} tissues. 

The fascicle-specific metrics estimated by Magic DIAMOND can be combined with tractography using their associated local orientations, as shown in Fig.~\ref{Figure_tracto} with the fascicle fractional anisotropy fFA mapped onto tracks associated to the corpus callosum, the corticospinal tract and the arcuate fasciculus. Outputted by our stratified bootstrap procedure, the median fFA along bundles is rather constant within the bulk of these bundles, except close to cortical grey matter and deep grey nuclei, within fascicle-crossing regions, and in high curvature areas (\textit{e.g.} high curvature area in the arcuate fasciculus, see panel (b) of Fig.~\ref{Figure_tracto}). The lower Med[fFA] in these regions is explained by the higher axonal dispersion which, as shown in Ref.~\onlinecite{Scherrer_aDIAMOND:2017}, is captured by an increased heterogeneity but also by a decreased fAD and increased fRD. These particular areas also appear bright in the IQR-colored bundles, reflecting a loss in precision on parameter estimation, in agreement with Fig.~\ref{Figure_orientations}.

Finally, Figs.~\ref{Figure_cost_function_CC_2} and \ref{Figure_cost_function_CS_2} show that the centrum-semiovale (CS) optimization landscape is flatter than the corpus-callosum (CC) optimization landscape in the $f$-$\lambda^\parallel$ and $f$-$\lambda^\perp$ projections, and is more peaked than the CC optimization landscape in the $\lambda^\parallel$-$\lambda^\perp$ projections. But we note that the optimization landscape exhibits no consistent change upon introduction of non-conventional b-tensors (LS and LP). This fact is reinforced by the three-dimensional illustration of the optimization landscape presented in Figs.~\ref{Figure_cost_function_CC_1} and \ref{Figure_cost_function_CS_1}: LL, LP and LS feature comparable separations between the minimum found by Magic DIAMOND and the global minimum of the sampled subspace.

\section{Conclusions}

In this work, we derive for the first time a general Laplace transform of the non-central matrix-variate Gamma distribution. This Laplace transform, valid for any arbitrary b-tensor, enables the design of the Magic DIAMOND model, \textit{i.e} the combination of the DIAMOND model, capable of separately characterizing crossing fascicles, with tensor-valued diffusion encoding. Our new theoretical approach aims at drawing from the complementary pieces of diffusion information yielded by tensor-valued diffusion encoding in order to refine the orientation and diffusional features of each intra-voxel fascicle estimated by DIAMOND. By design, it therefore offers new possibilities to quantify the optimization of tensor-valued encoded acquisition schemes \citep{Bates_ISMRM:2019}. 

Our \textit{in vivo} evaluations suggest that introducing non-conventional b-tensors (either planar or spherical) leads to a lower estimated free-water signal fraction in the white matter (Fig.~\ref{Figure_fFW}). This change is in agreement with the understanding that there is none or very little freely diffusing water in the densely packed white matter when using typical clinically feasible diffusion times \citep{Dhital:2018}, which in turn supports the fact that non-conventional b-tensors sample new information about the underlying tissue microstructure and allow improved separation of isotropic and anisotropic diffusion components. Importantly, our statistical evaluation \textit{via} stratified bootstrap shows that spherical encoding consistently lowers the precision on parameter estimation. This was verified in both our voxel-based evaluation throughout white matter (Figs.~\ref{Figure_orientations}-\ref{Figure_fFW}-\ref{Figure_fAD}-\ref{Figure_fRD}) and in our fixel-based evaluation in which we colored each tract streamline with the IQR of the corresponding estimated compartment (Fig.~\ref{Figure_tracto}). First, this is likely explained by the intrinsic lack of information linking the fascicles' orientations to their diffusional features when using spherical encoding (see Eq.~\eqref{Eq_Magic-DIAMOND_spherical}). Second, the inherently low signal-to-noise ratio (SNR) of spherical encoding may also be involved in lowering precision. In a conventional linear acquisition, the diffusion-attenuated signal is low when the diffusion gradient orientation is aligned with a fascicle orientation. Because spherical encoding samples all orientations, a decrease of spherical signal occurs for any fascicle orientation, leading to an SNR that is always lower than that with linear encoding. However, the SNR depends on the acquisition scheme and scanning hardware, so that the difference in SNR between distinct tensor-valued acquisitions may be mitigated in other acquisition setups. As a result, we found that at constant scan duration time, the isotropic information brought by spherical encoding has too low SNR and not enough information about the fascicles themselves to improve the estimation of fascicle-specific diffusivities and orientation of a multi-fascicle model without increasing uncertainty. In contrast, the introduction of planar encoding did not lower precision on parameter estimation. However, we surprisingly did not observe a clear advantage of our linear-planar dataset (LP) over our purely linear dataset (LL). Indeed, not only did it not reduce angular uncertainty on the fascicles' orientations (Fig.~\ref{Figure_orientations}), unlike what could have been expected \citep{Cottaar:2019}, it also did not provide a more constraining optimization landscape (Figs.~\ref{Figure_cost_function_CC_1}-\ref{Figure_cost_function_CS_1}-\ref{Figure_cost_function_CC_2}-\ref{Figure_cost_function_CS_2}).

Nonetheless, combining non-trivial diffusion encodings with the multi-fascicle DIAMOND model should not be discarded right away. Indeed, potential improvements upon introducing tensor-valued diffusion encoding may be yielded by the following families of solutions. 
First, from a data-acquisition standpoint, one could for instance try to obtain higher SNRs and/or higher b-values, by using higher gradient amplitudes and shortening TE as in Ref.~\onlinecite{Szczepankiewicz_DIVIDE:2019}, in order to further differentiate linear and planar pieces of diffusion information~\citep{Jespersen:2012}. Alternatively, the acquisition protocol could be further optimized by maximizing correlations between acquisition dimensions so that to bring tighter constraints to the Magic-DIAMOND minimization. Combining at least two diffusion encodings might also simply make Magic DIAMOND's parameter estimation more robust to data undersampling, an effect that has already been observed in the past~\citep{Szczepankiewicz_Jeurissen_ISMRM:2018}. 
Second, from a data-processing point of view, phase correction strategies \citep{Eichner:2015,Pizzolato:2020} may be practical given the inherently lower SNR of planar and spherical encodings compared to linear encoding. 
Third, the current design of the DIAMOND model may be too constrained to take advantage of the power of tensor-valued diffusion encoding. Such limitation was previously noticed in Ref.~\onlinecite{Lampinen_CODIVIDE:2017} when using spherical encoding within the original NODDI model \citep{Zhang_NODDI:2012}. The importance of DIAMOND's constraints could be investigated by opening up the model so that it becomes degenerate for LL but not for LP or LS.
As the mathematics of the Magic DIAMOND model are now laid down, we invite the diffusion MRI community to use them and challenge or reproduce our conclusions.

\section*{Acknowledgments}
The authors would like to thank the "Axe Imagerie M\'{e}dicale du Centre de recherche du CHUS" and the "Centre d'imagerie m\'{e}dicale de l'Universit\'{e} de Sherbrooke" for sponsored image acquisition time. The study was also supported by the Neuroinformatics Chaire of the Universit\'{e} de Sherbrooke. We also thank the Quebec bio-imaging network (QBIN) for partial funding of first author A. Reymbaut. 


\end{document}